\documentclass[preprint2]{aastex631}

\newcommand\NBBH{79}
\newcommand\alphap{1\%}
\newcommand\vvmax{$\mathrm{V/V_\mathrm{max}}$}
\newcommand\nnmax{$\mathrm{N/N_\mathrm{max}}$}

\usepackage{graphicx}
\usepackage{booktabs}

\received{November, 2021}
\revised{April, 2022}
\accepted{April, 2022}

\shorttitle{Are BBH mergers and LGRBs drawn from the same BH population?}
\shortauthors{Arcier et Atteia}

\graphicspath{{./}}

\begin{document}

\title{Are binary black hole mergers and long $\gamma$-ray bursts drawn from the same black hole population?}

\correspondingauthor{Benjamin Arcier}
\email{Benjamin.Arcier@irap.omp.eu}

\author{Benjamin Arcier}
\affiliation{IRAP, Universit\'e de Toulouse, CNES, CNRS, UPS, Toulouse, France}

\author{Jean-Luc Atteia}
\affiliation{IRAP, Universit\'e de Toulouse, CNES, CNRS, UPS, Toulouse, France}

\begin{abstract}

This paper compares the population of BBH mergers detected by LIGO/Virgo with selected long GRB world models convolved with a delay function (LGRBs are used as a tracer of stellar mass BH formation). The comparison involves the redshift distribution and the fraction of LGRBs required to produce the local rate of BBH mergers.

We find that BBH mergers and LGRBs cannot have the same formation history, unless BBHs mergers have a long coalescence time of several Gyr. This would imply that BHs born during the peak of long GRB formation at redshift $z\approx2-3$ merge within the horizon of current GW interferometers. We also show that LGRBs are more numerous than BBH mergers, so that most of them do not end their lives in BBH mergers.

We interpret these results as an indication that BBH mergers and LGRBs constitute two distinct populations of stellar mass BHs, with LGRBs being more frequent than BBH mergers.
We speculate that the descendants of LGRBs may resemble galactic High-Mass X-Ray Binaries more than BBH mergers. We finally discuss the possible existence of a sub-population of fast-spinning LGRBs descendants among BBH mergers, showing that this population, if it exists, is expected to become dominant beyond redshift \mbox{$z \approx 1$}, leading to a change in the observed properties of BBH mergers.

\end{abstract}


\section{Introduction}
\label{sec:introduction}

Black hole (BH) astrophysics is a rapidly evolving field, especially thanks to the discovery of dozens of binary black hole mergers by the LIGO Scientific Collaboration and Virgo Collaboration (LVC). Despite the apparent simplicity of black holes as astrophysical objects, a global vision of their origin and evolution is still missing as well as a full understanding of their role in the evolution of the universe and in the formation of its structures. In such a situation, the comparison of the properties of various BH populations may shed light on their common or diverse origins.
We focus here on two populations of stellar mass black holes, which are observed in different contexts: long gamma-ray bursts (long GRBs or LGRBs) and the mergers of stellar mass black holes (BBH mergers).

Long GRBs are the brightest explosions detectable in the electromagnetic domain \cite[e.g.][]{Vedrenne2009, Atteia2017}. They occur when the core of a massive star collapses into a black hole or an hypermassive magnetar, subsequently ejecting a transient relativistic jet in our direction. The association of LGRBs with stellar collapse is confirmed by the detection of broad-line core-collapse supernovae of type Ibc following several nearby LGRBs (z $\leq 0.3$) after a few days \citep{Galama1998_980425, Tagliaferri2006_031203}. However, only a small fraction of core collapse supernovae produce LGRBs.

This fraction has been evaluated by \citet{Soderberg2006a}, based on volumetric rates of about $9000\ \mathrm{Gpc^{-3}\, yr^{-1}}$  for supernovae Ibc and $f_\mathrm{b} \times 1\ \mathrm{Gpc^{-3}yr^{-1}}$ for classical LGRBs \citep[e.g.][]{Wanderman2010, Palmerio2021}, where $f_\mathrm{b}$ is the GRB beaming factor. It is difficult to evaluate the true rate of GRBs, due to large uncertainties on the measurements of the beaming angle of GRB jets (thus on $f_\mathrm{b}$). With typical values of $f_\mathrm{b}$ in the range $75-250$ (corresponding to jet opening angle in the range $5$°$-9$°), the classical LGRBs represent from 1 to 3\% of SNIbc.
This is compatible with another study of \citet{Soderberg2006b}, which shows that less than 10\% of SNIbc are associated with a successful GRB, based on the radio followup of 68 local supernovae of Type Ibc.
We note that this fraction concerns only classical LGRBs, excluding low-luminosity GRBs, which have also been shown to be associated with SNIbc, or choked GRBs, whose jet does not pierce the star's envelope.
The rate of low-luminosity GRBs is estimated to be $\sim100-1000\ \mathrm{Gpc^{-3}~yr^{-1}}$ \citep{Soderberg2006a, Liang2007}, comparable or a few times higher than the rate of classical LGRBs. The rate of choked GRBs has not been measured.

The conditions for GRB production by dying massive stars is far from being fully elucidated.
\citet{MacFadyen1999} emphasized the key role of the angular momentum of the stellar core, which has to be sufficiently large to permit the survival of a massive accretion disk when the BH forms. Before the discovery of BBH mergers with low effective spins by the LVC, it was commonly assumed that GRB progenitors could be single stars with low metallicity, whose weak stellar winds carry away a small fraction of the star angular momentum or stars in binary systems, which are sufficiently close to be tidally locked and keep a high angular momentum \cite[e.g.][]{Woosley2006, Levan2016, Chrimes2020}. 
Except for a few events with an associated supernova, we have only indirect information about the progenitors of LGRBs, because the information is mediated by the relativistic jet, whose properties are not directly connected with the nature of the progenitor. The population of long GRB progenitors can nevertheless be constrained via statistical studies relying on redshift measurements and the nature of their host galaxies. This has led to the construction of long GRB world models, which describe the history of long GRB formation and are compatible with their observed redshift distribution. \citep{Salvaterra2012, Lien2014, Palmerio2021}

Contrary to LGRBs, BBH mergers are mostly silent in the electromagnetic domain (see however \citealt{Farris2010, Bartos2017, McKernan2020} for discussions about the possible emission of electromagnetic transients by BBHs lying in a dense gaseous environment). These cataclysmic events, which have been discovered by the LVC in 2015 \citep{Abbott2016}, release most of their energy in gravitational waves (GW).
The origin of these BBH mergers is highly debated, and the variety of systems discovered by the LVC probably calls for multiple populations \citep{Zevin2021, Wang2021}, however see \citealt{Bavera2020}. There is nevertheless a broad consensus that most of the mergers detected by the LVC are of astrophysical origin, even if a marginal contribution from primordial black holes remains possible \citep{Luca2021}. BBH mergers of astrophysical origin result from the evolution of massive stars, which can be in isolated binary systems or in a dense environment prone to dynamical interactions. 

Considering the different biases affecting the detection of BBH mergers and LGRBs, the comparison of these two populations is likely to shed new light on the origin and evolution of stellar mass black holes of astrophysical origin.

In this article, we compare basic statistics concerning these two populations. First a \vvmax\ test is used to compare the redshift distribution of BBH mergers with various delayed long GRB models, where long GRB models are used as proxies for the formation history of stellar mass black holes. This comparison, which takes into account the shape of the redshift evolution independent of the normalization of the two populations,  leads us to identify several models compatible which can reproduce the redshift distribution of BBH mergers detected by LVC.
In a second step, we compare the rate of BBH mergers measured by LVC with the rate of long GRB. This comparison makes no assumption of a physical connection between the two populations, it only uses the rate of long GRBs as a useful reference for the comparison of populations with different redshift evolution.
We find that the progenitors of BBH mergers are rare compared with GRBs.
For these two studies the BBH mergers are considered as a single population.\\
These findings are then briefly discussed in the context of models developed to explain the BBH mergers and the GRBs. Emphasizing the potential role of tidal spin-up for both types of sources, we show that present data allow the existence of a minority of fast-spinning BBH mergers that quickly follow the production of a LGRBs.

This study has been made possible by the publication of the GWTC-2 and GWTC-3 catalogs obtained with the O3 observing run \citep{GWTC2, GWTC3}. These catalogs contain \NBBH\ BBH mergers with redshifts ranging from \mbox{z = 0.05} to \mbox{z = 0.82} with a median of \mbox{z = 0.3}), enabling the comparison of the redshift distributions of BBH mergers with GRBs over the last 7 Gyrs. 
It follows a first work from \cite{Atteia2018}, based on the GWTC-1 catalog \citep{GWTC1} and thus limited to the rates comparison of BBHs and LGRBs.

This paper is organized as follows: in the next two sections we compare the redshift distribution (Sect. \ref{sec:distriz}) and the volumetric rate (Sect. \ref{sec:rate}) of BBH mergers and LGRBs.
Then, in Sect. \ref{sec:discussion}, we briefly discuss the implication of these results in the context of stellar evolution models developed to explain LGRBs and BBH mergers.

In all this paper, we use a flat $\Lambda\mathrm{CDM}$ cosmological model with the parameters measured by the Planck Collaboration: $H_\mathrm{0} = 67.4~\mathrm{km~s^{-1}}~\mathrm{Mpc}^{-1}$ and $\Omega_\mathrm{m} = 0.315$ \citep{PlanckCollaboration2018}.

\section{Redshift distribution}
\label{sec:distriz}
The fast growing number of BBH mergers detected by the LVC provides some indications on their redshift distribution, as discussed in \citet{Abbott2021} and \citet{GWTC3_population}.
Using a generalization of the \vvmax\ test developed by \cite{Schmidt1968},
we compare here the redshift distribution of \NBBH\ BBH mergers detected with high confidence by the LVC with three recent models of classical LGRBs and one star formation history (SFH) model.

\subsection{Methodology}
\label{sub:method}

We use a test called \nnmax , which is a variant of the \vvmax\ test developed by \cite{Schmidt1968}. These tests are used to assess the compatibility of the volumetric distribution of observed sources with a model. In the \vvmax\ test, the model is used to compute, for each source, the ratio of the volume enclosed by the source to the maximum volume of detection for this source. If the model is correct, the source is randomly chosen among all visible sources and the ratio \vvmax \ follows a uniform distribution $\mathcal{U}(0,1)$. In its simplest form, the \vvmax\ test considers sources without cosmological evolution. For sources with cosmological evolution, the test can be adapted by considering the ratio of the predicted number of sources within the volume enclosed by a source to the number of sources within the maximum volume of detection for this source.
When the model is completely defined and the volume of detection is correctly calculated, the \vvmax\ and \nnmax\ tests are hardly affected by selection effects, because each source is placed within its own volume of detection.

The volume of detection of each BBH merger is computed via the determination of its \textit{horizon} $z_\mathrm{h,i}$, which is the maximum redshift at which this merger could be detected (see below). Knowing the merger event redshift $z_\mathrm{i}$ and its horizon redshift $z_\mathrm{h,i}$, we can compute $\mathrm{N_i}$, the number of sources closer than $z_\mathrm{i}$, and $\mathrm{N_\mathrm{max,i}}$, the number of sources closer than the horizon $z_\mathrm{h,i}$, for various source models. A model is acceptable if the ratio $\mathrm{N_i/N_\mathrm{max,i}}$ follows a uniform distribution $\mathcal{U}(0,1)$. The agreement of the distribution of $\mathrm{N_i/N_\mathrm{max,i}}$ with $\mathcal{U}(0,1)$ is evaluated with a Kolmogorov-Smirnov test (hereafter, KS test). This comparison does not take into account the normalization of the GRB and BBH merger rates, which is discussed in Sect. \ref{sec:rate}.

The method used to determine the horizon redshift is based on the horizon calculation method proposed in \cite{Chen2021} and the open-source code available at \url{https://github.com/hsinyuc/distancetool}.
The equation to compute the SNR evolution with the redshift is taken from \cite{Chen2021}:

\begin{equation}
\mathrm{SNR_\mathrm{Ch21}} = \sqrt{4\int_{f_\mathrm{min}}^{f_\mathrm{max}}\frac{|h^{+}(f)|^2}{S_\mathrm{h}(f)}\mathrm{d}f}
\label{equ:SNR_chen}
\end{equation}

where $h^{+}(f)$ is the plus polarization of the gravitational-wave merger and $S_\mathrm{h}(f)$ is the power spectral density of the interferometer.
The calculation of $h^{+}(f)$ depends on the binary properties and its distance relatively to the observer, in addition to the masses $m_1$ and $m_2$ of the BBH merger components. This plus polarization is obtained using the \texttt{lalsimulation} Python package with the waveform \textit{IMRPhenomD} (as for the online calculator provided by \citealt{Chen2021}, since $m_1$ or $m_2 \geq 3\ \mathrm{M_\odot}$).
For the term $S_\mathrm{h}(f)$, we took the same strain noise curve for LIGO Handford and LIGO Livingston, for a given observing run, already included in the Github repository from \cite{Chen2021}.

Contrary to \cite{Chen2021}, who define the horizon as the highest redshift at which a BBH merger could be detected with the best antenna pattern (considering that the BBH system is optimally oriented relative to the GW detectors and located in a portion of the sky where the interferometers performances are maximum), we define the horizon as the maximum redshift at which a BBH merger would be detected, \textit{under the conditions of its detection}.
Since the antenna pattern does not depend upon distance and cosmology, it is possible to normalize the SNR dependence on redshift with the actual signal-to-noise ratio $\mathrm{SNR_0}$ obtained for a given merger detected at redshift $z_0$. We obtain the SNR dependence on redshift corrected by the antenna pattern, allowing to compute the horizon redshift for each detected merger.

\begin{equation}
\mathrm{SNR}(z, m_1, m_2) = \frac{\mathrm{SNR}_\mathrm{Ch21}(z, m_1, m_2)}{\mathrm{SNR}_\mathrm{Ch21}(z_0, m_1, m_2)}\times \mathrm{SNR_0} 
\end{equation}

Taking a threshold $\mathrm{SNR_{lim}}$, the horizon redshift is the variable $z_\mathrm{h,i}$ solving the equation:

\begin{equation}
\label{equ:z_h}
\mathrm{SNR}(z_\mathrm{h,i}, m_1, m_2) = \mathrm{SNR_{lim}}
\end{equation}

\begin{figure*}
    \centering
    \resizebox{\hsize}{!}{\includegraphics{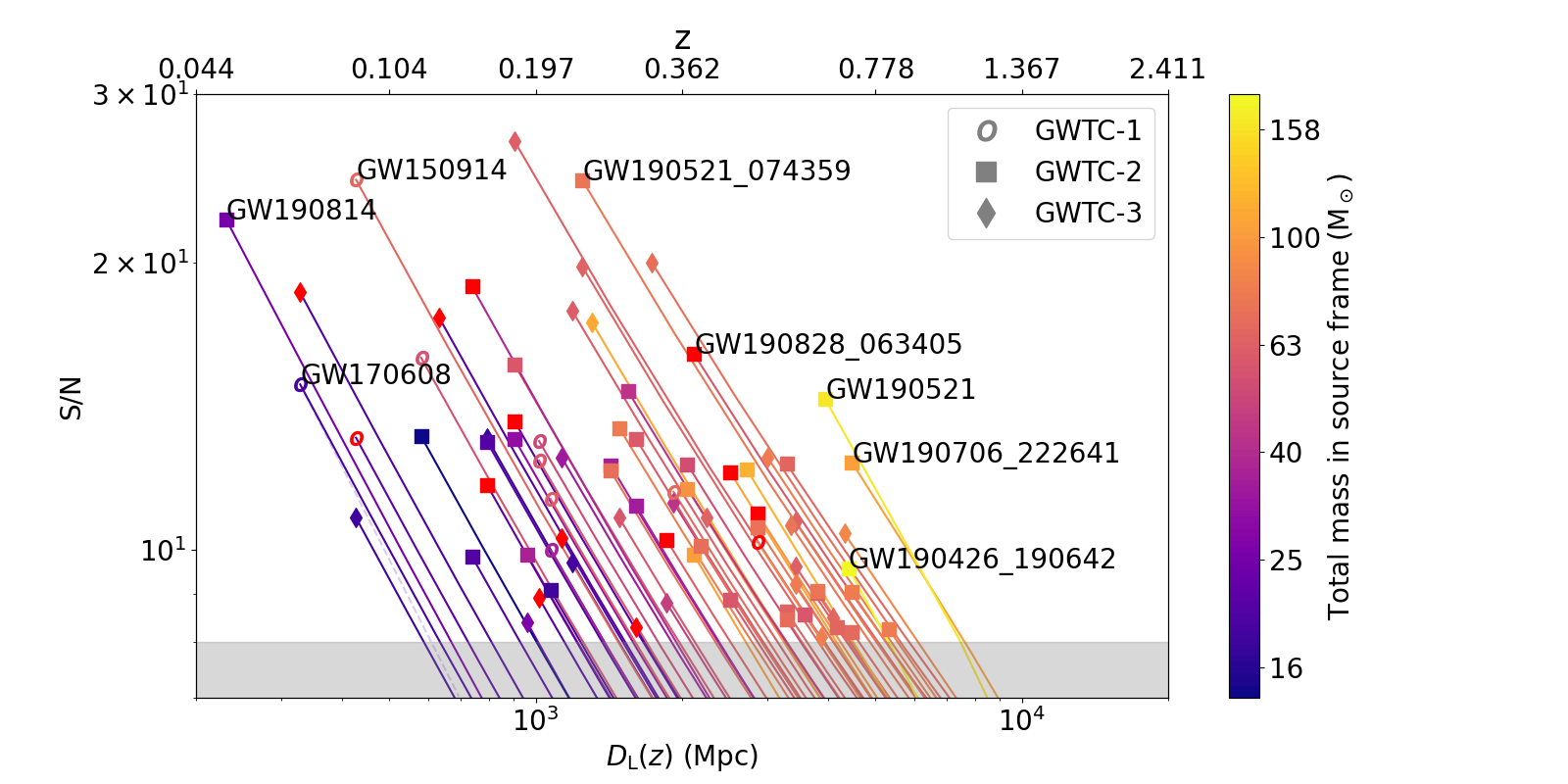}}
    \caption{SNR dependence on redshift for the BBH mergers observed during O1, O2, O3a and O3b. The color coding represents the total mass of the binary in the observer frame, while the marker style represents the catalog in where the events are listed. The dashed line for GW~170608 illustrates a SNR dependency inversely proportional to the luminosity distance.}
    \label{fig:zh_GWTC}
\end{figure*}

The SNR, mass and detected redshift are taken from the Gravitational Wave Open Science Center (GWOSC) event list\footnote{\url{https://www.gw-openscience.org/eventapi/html/GWTC/}}, created from the GWTC catalogs \citep{GWTC1, GWTC2, GWTC2_1, GWTC3}.
In the following, we consider $\mathrm{SNR_{lim}} = 8$ in agreement with the smallest SNR observed for BBHs mergers \citep{GWTC1, GWTC2}, and with the theoretical sensitivities of LIGO/Virgo \citep{prospectsGW_2018}. Recent merger observations from the deep extended catalog from O3a \citep{GWTC2_1} and from GWTC-3 \citep{GWTC3} have however detected events with a network matched filter SNR below this limit. These events are not considered in this study. Events with mass $m_1$ and/or $m_2$ below $2.4\ \mathrm{M_\odot}$ have also been removed, to prevent the contamination from NS-NS or NS-BH merger events.

Table \ref{tab:z_tab} gives the horizons of the \NBBH\ events used in this article, computed with this method. Figure \ref{fig:zh_GWTC} illustrates the redshift dependence and the impact of the shape of the aLIGO power-spectral density (which is not exactly flat), which causes the departure from the inverse luminosity distance law (shown with a faint dashed line for GW~170608). 
Mergers in GWTC-1, GWTC-2 and GWTC-3 have a broad range of horizons, from \mbox{z = 0.12} to \mbox{z = 1.10} with a median \mbox{z = 0.50}. This is the reflection of their mass range, which spans over 1 decade. This broad range of horizons implies very different detection volumes for massive ($\sim 160~M_\sun$) and less massive BBH mergers ($\leq 20~M_\sun$), and a mass distribution of detected mergers which is strongly biased in favor of massive BBHs. By construction the \nnmax\ test takes into account this bias by considering the individual contribution of each BBH merger within its own volume of detection.

The \nnmax\ test requires computing $\mathrm N_\mathrm{i}$, the number of mergers expected up to redshift  $z_\mathrm{i}$ and $\mathrm N_\mathrm{max,i}$, the number of mergers expected up to redshift $z_\mathrm{h,i}$. This is based on a world model of source population that specifies $\rho(z)$, the evolution of the source density rate with redshift. $\mathrm N_\mathrm{i}$ is computed as follows:

\begin{equation}
\label{equ:nz}
    \mathrm{N_i} = \int_0^{z_\mathrm{i}} \rho(z)\ 
    \frac{\mathrm{d}V(z)}{\mathrm{d}z}\ \frac{1}{1+z}\ \mathrm{d}z
\end{equation}

Where $\rho(z)$ is the BBHs merger density rate (in $\mathrm{Gpc^{-3}\ yr^{-1}}$), $\mathrm{d}V(z)$ is the differential co-moving volume, and the term $1/(1+z)$ accounts for the impact of time dilation when we measure a \textit{rate} of events \citep{Mao1992}. 

\begin{figure*}
\gridline{\fig{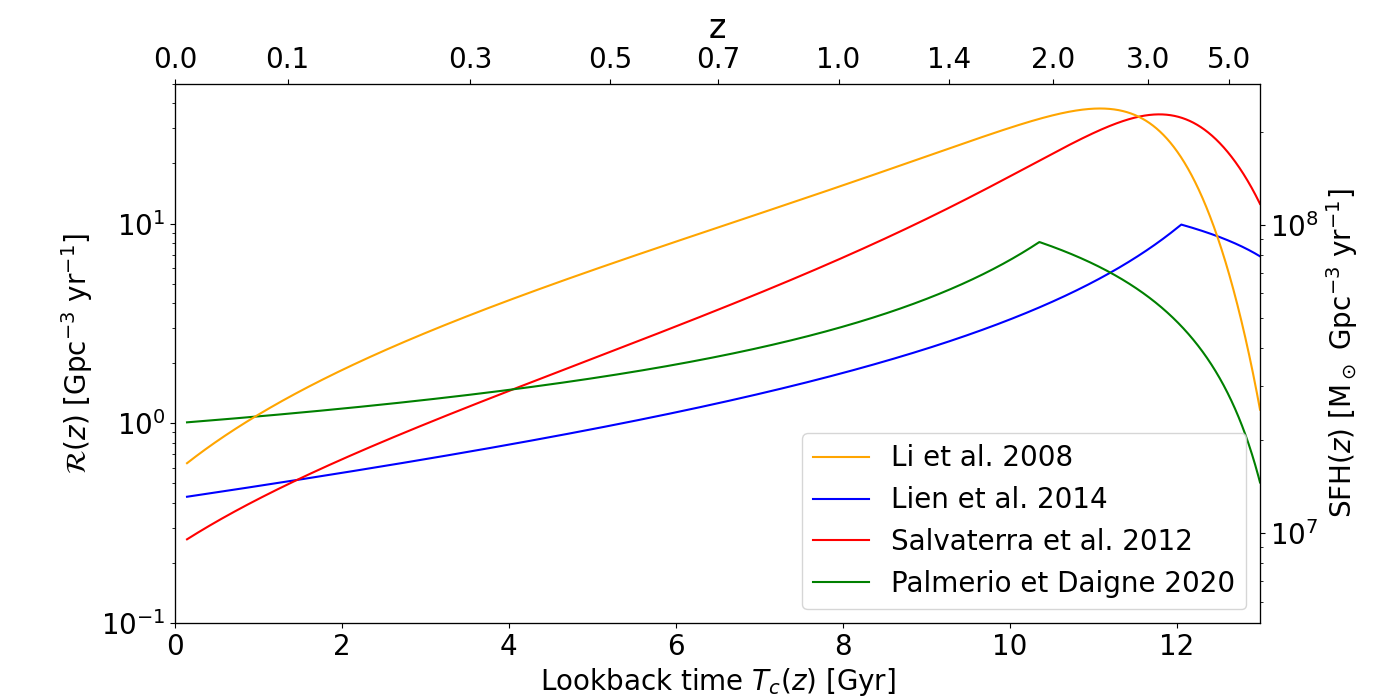}{0.45\textwidth}{\label{fig:Rgrb}(a) GRB density rate for the three GRB world models (blue, red and green) and the SFH (orange).}
\fig{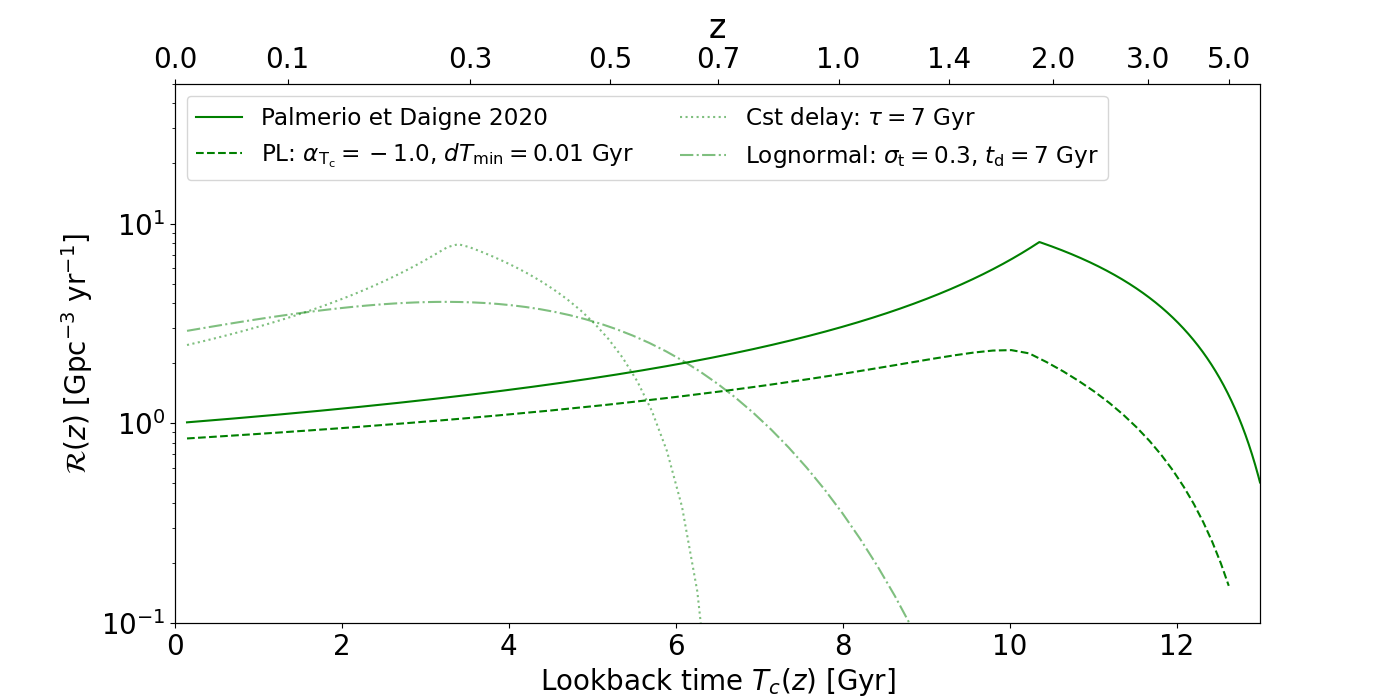}{0.45\textwidth}{\label{fig:Rgrb_d}(b) GRB density rate obtained from \citeauthor{Palmerio2021} convolved with three shapes of the delay function: dotted for the power-law shape, dashed for the constant delay shape and dash-dotted for a lognormal shape.}}
\gridline{\fig{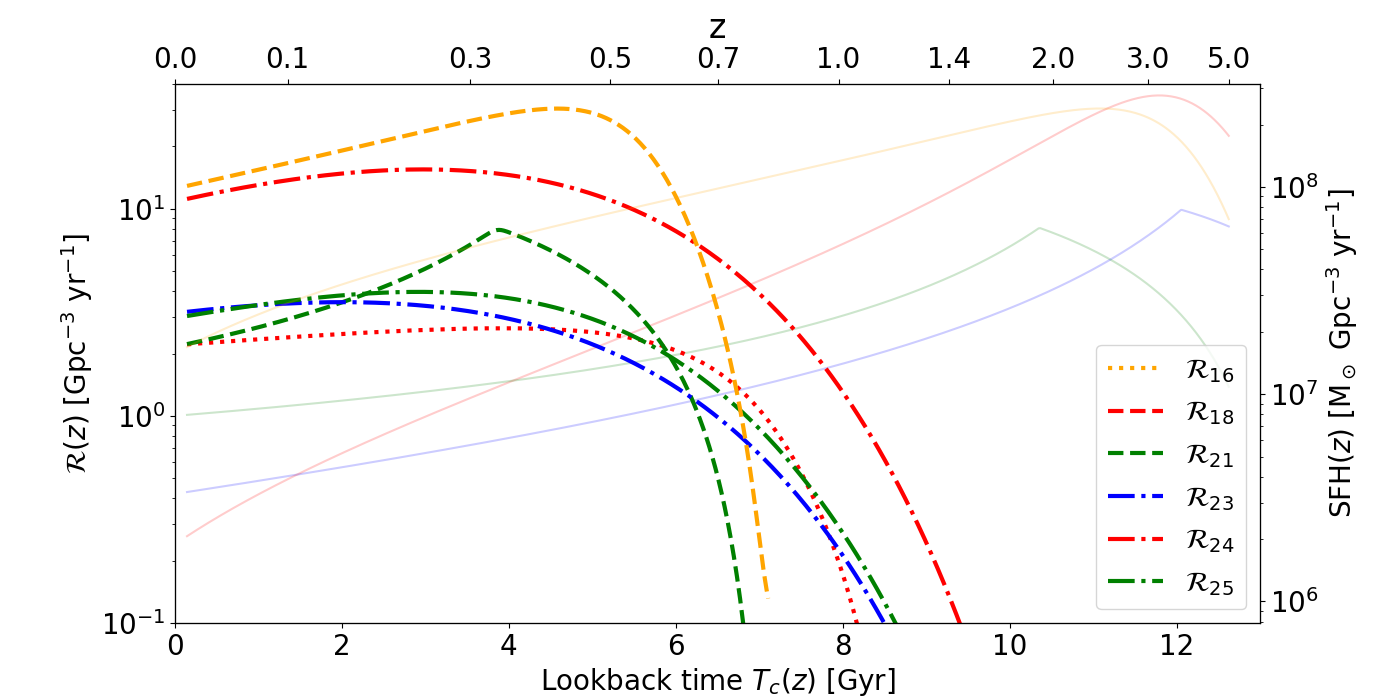}{0.45\textwidth}{\label{fig:Rgrb_best}(c) Some models with a p-value above $0.1$ (dotted and dashed lines). The corresponding models can be found in Table \ref{tab:ks_results}. The faint full line plots are the GRB/SFH models before the delay function convolution.}}
\caption{\label{fig:Rgrb_z} GRB density rate models and SFH evolution with redshift, including several time delay functions.}

\end{figure*}

We compare here the observed distribution of BBH mergers with several density rate models: a constant rate model, a model of SFH and three GRB world models which all reproduce the observed properties of GRBs detected by the \textit{Swift} and \textit{Fermi} missions. These three models and the SFH are illustrated in Fig. \ref{fig:Rgrb_z}a.

The first GRB model is proposed by \cite{Salvaterra2012}:
\begin{equation}
    \mathcal{R}_\mathrm{GRB}^\mathrm{S}(z) = \mathcal{R}_\mathrm{0}^\mathrm{S}\ (1+z)^\mathrm{\delta_n}\ \mathrm{\Sigma}_\mathrm{SFH}^\mathrm{Li}(z)
\end{equation}
based on the cosmic SFH rate from \cite{Li2008}:
\begin{equation}
    \mathrm{\Sigma}_\mathrm{SFH}^\mathrm{Li}(z) = \frac{a+bz}{1+(z/c)^d}
\end{equation}
with $\mathcal{R}_\mathrm{0}^\mathrm{S} = 0.24\ \mathrm{Gpc}^{-3}\ \mathrm{yr}^{-1}$, $\delta_n=1.7$, $a=0.0158$, $b=0.118$, $c=3.23$ and $d=4.66$.\\

The second is proposed by \cite{Lien2014}:
\begin{equation}
    \mathcal{R}_\mathrm{GRB}^\mathrm{L}(z) =  \mathcal{R}_\mathrm{0}^\mathrm{S}\ \left\{
    \begin{array}{ll}
        (1+z)^\mathrm{n_1} & \mbox{if } z < \mathrm{z_1} \\
        (1+\mathrm{z_1})^\mathrm{(n_1-n_2)} (1+z)^\mathrm{n_2} & \mbox{if } z \geq \mathrm{z_1}
    \end{array}
\right.
\end{equation}
with $\mathcal{R}_\mathrm{0}^\mathrm{L} = 0.42\ \mathrm{Gpc}^{-3}\ \mathrm{yr}^{-1}$, $n_1=2.07$, $n_2=-0.7$ and $\mathrm{z_1}=3.6$.\\

The third is proposed by \cite{Palmerio2021}:

\begin{equation}
    \mathcal{R}_\mathrm{GRB}^\mathrm{P}(z) = \mathcal{R}_\mathrm{0}^\mathrm{P}\ \left\{
    \begin{array}{ll}
        e^{az} & \mbox{if } z < \mathrm{z_m} \\
        e^{(a-b)\mathrm{z_m}} e^{bz} & \mbox{if } z \geq \mathrm{z_m}
    \end{array}
\right.
\end{equation}
with $\mathcal{R}_\mathrm{0}^\mathrm{P} = 1.0\ \mathrm{Gpc}^{-3}\ \mathrm{yr}^{-1}$, $a=1.1$, $b=-0.57$ and $\mathrm{z_m}=1.9$.\\

Since the population of BBHs mergers is probably not directly tracing the GRB population, we also consider delayed models with a delay function between the GRB emission (when the BH is created) and the transient GW emission (when the BH merges with its companion). This approach is similar to the studies that compare the rates of binary neutron star mergers and short GRBs \citep{Nakar2007, Wanderman2015}. Considering a time delay probability density function $f(\tau)$ and integrating over all possible time delays, the calculation of the BBH merger rate at redshift $z_0$ becomes:

\begin{equation}
\label{equ:Rbbhs_z}
    \rho(z_0) \propto \int_{z_0}^{\infty} \mathcal{R}_\mathrm{GRB}(z)\ f(T_c(z) - T_c(z_0))\ \frac{\mathrm{d}T_c}{\mathrm{d}z}\ \mathrm{d}z
\end{equation}
where $T_c(z)$ is the lookback time at redshift $z$, and $\tau = T_c(z) - T_c(z_0)$ is the time delay in Gyr between a GRB at redshift $z$ and a merger produced at redshift $z_0 < z$. In practice, a limiting redshift $z_\mathrm{lim} = 20$ has been chosen for the calculation of the integral, and we have verified that changing this value from $20$ to $100$ has no significant impact on the results. The delay function $f(\tau)$ can have several shapes, and in the following we consider three of them: a power-law, a lognormal distribution, and a constant delay. 

The power-law has two parameters: a negative slope $\alpha$ (for example $\alpha=-1$ in \citealt{Belczynski2016}) and a minimum merging time $dT_\mathrm{min}$. It is described by:

\begin{equation}
    f(\tau) = \left\{
    \begin{array}{ll}
        0 & \tau \leq dT_\mathrm{min} \\
        \tau^\alpha & \tau > dT_\mathrm{min}
    \end{array}
\right.
\end{equation}

The lognormal distribution of width $\sigma_\mathrm t$ centered around a time delay $t_\mathrm d$ is described by:

\begin{equation}
    f(\tau) = \frac{1}{\tau\sigma_\mathrm{t}\sqrt{2\pi}} \exp \left(-\frac{\ln(\tau/t_\mathrm{d})^2}{2\sigma_\mathrm{t}^2}\right) 
\end{equation}

The constant delay is modeled by setting a small dispersion of the lognormal distribution (typically $\sigma_\mathrm{t} = 0.01$). This is how the constant delay GRB density rate presented on Fig. \ref{fig:Rgrb_z}b has been created.
From the GWTC-1, GWTC-2 and GWTC-3 catalogs, the maximum redshift observed for a BBHs merger is $z=0.71$, which represents a lookback time of $\sim 6.5$ Gyr, with the cosmology used in this paper. For this reason, we have not considered constant delay models with a delay greater than $7$ Gyr, which would lead to produce the most distant BBH mergers before the Big Bang. The differential of the lookback time at redshift $z$ is calculated using the definition of the lookback time in \citealt{Condon2018}:

\begin{equation}
    \frac{\mathrm{d}T_\mathrm{c}}{\mathrm{d}z}(z) = \frac{1}{H(z)(1+z)}
\end{equation}

Fig. \ref{fig:Rgrb_z}b illustrates the impact of these different delay functions on the GRB model of \citealt{Palmerio2021}. 

\subsection{Results}
\label{sub:results}

\begin{figure}
    \centering
    \resizebox{\hsize}{!}{\includegraphics{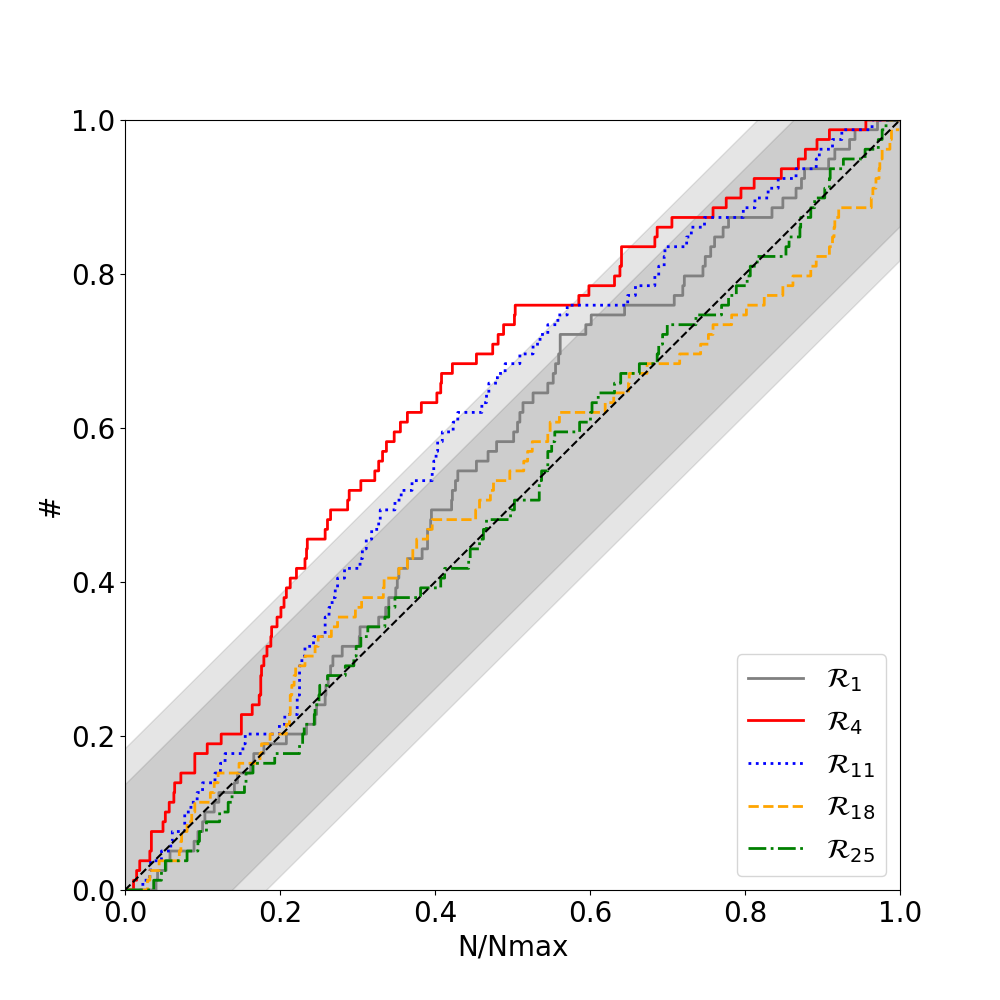}}
    \caption{Comparison of $\mathrm{N/N_\mathrm{max}}$ cumulative distributions with the expected $\mathcal{U}(0,1)$ distribution. The shaded areas represent the \alphap\ and 10\% acceptance regions for  the KS test. $\mathcal{R}_1$ represents the constant model, $\mathcal{R}_4$ the GRB model from \cite{Salvaterra2012}, $\mathcal{R}_{11}$ the GRB model from \cite{Lien2014} convolved with a PL delay function (with $\alpha=-2$ and $dT_\mathrm{min}=0.01\ \mathrm{Gyr}$), $\mathcal{R}_{18}$ the SFH from \cite{Li2008} with a constant delay of $6.5\ \mathrm{Gyr}$ and $\mathcal{R}_{25}$ the GRB model from \cite{Palmerio2021} convolved with a lognormal delay function (with $\mathrm{t_d}=7.3\ \mathrm{Gyr}$ and $\sigma_\mathrm{t} = 0.3$).
    }
    \label{fig:models_comparison}
\end{figure}

Using the \nnmax\ test defined in the previous section, we now compare the redshift distribution of BBH mergers in the GWTC catalogs with various delayed GRB models. A model is considered acceptable if the KS test gives a p-value larger than \alphap\ and favored if the p-value exceeds $10$\%. This is illustrated in Fig. \ref{fig:models_comparison}, which compares the cumulative \nnmax\ distributions of various models with the data. The light grey zone indicates the 99\% confidence region, while the dark grey zone the 90\% confidence interval: a model is accepted at a given confidence level if its \nnmax\ cumulative distribution function lies entirely inside the grey zone.

\begin{table*}
\caption{Results of the Kolmogorov-Smirnov test for various population models (see text). The p-values of favored models are indicated in boldface.}
\label{tab:ks_results}
\resizebox{\hsize}{!}{
\begin{tabular}{llllrl}
\toprule
                        \# &            Population model &                                    Delay parameters &     p-value &  $\left< \mathrm{N/N_{max}}\right>$ &                                                                Fig. \\
\midrule
 $\mathcal{R}_\mathrm{1}$ &  Constant &  ... &  \it{0.035} & 0.452 &  \ref{fig:models_comparison} \\
 $\mathcal{R}_\mathrm{2}$ &  SFH: \cite{Li2008} &  ... &  0.001 & 0.389 &  \ref{fig:Rgrb}, \ref{fig:models_comparison} \\
 $\mathcal{R}_\mathrm{3}$ &  GRB: \cite{Lien2014} &  ... &  0.005 & 0.406 &  \ref{fig:Rgrb} \\
 $\mathcal{R}_\mathrm{4}$ &  GRB: \cite{Salvaterra2012} &  ... &  0.000 & 0.356 &  \ref{fig:Rgrb} \\
 $\mathcal{R}_\mathrm{5}$ &  GRB: \cite{Palmerio2021} &  ... &  \it{0.014} & 0.417 &  \ref{fig:Rgrb}, \ref{fig:Rgrb_d} \\
 $\mathcal{R}_\mathrm{6}$ &  SFH: \cite{Li2008} &  PL, $dT_\mathrm{min}=0.01$ Gyr, $\alpha=-1.0$ &  0.006 & 0.412 &  ... \\
 $\mathcal{R}_\mathrm{7}$ &  GRB: \cite{Lien2014} &  PL, $dT_\mathrm{min}=0.01$ Gyr, $\alpha=-1.0$ &  \it{0.012} & 0.419 &  \ref{fig:models_comparison} \\
 $\mathcal{R}_\mathrm{8}$ &  GRB: \cite{Salvaterra2012} &  PL, $dT_\mathrm{min}=0.01$ Gyr, $\alpha=-1.0$ &  0.003 & 0.401 &  ... \\
 $\mathcal{R}_\mathrm{9}$ &  GRB: \cite{Palmerio2021} &  PL, $dT_\mathrm{min}=0.01$ Gyr, $\alpha=-1.0$ &  \it{0.017} & 0.425 &  \ref{fig:Rgrb_d} \\
 $\mathcal{R}_\mathrm{10}$ &  SFH: \cite{Li2008} &  PL, $dT_\mathrm{min}=0.01$ Gyr, $\alpha=-2.0$ &  0.001 & 0.390 &  ... \\
 $\mathcal{R}_\mathrm{11}$ &  GRB: \cite{Lien2014} &  PL, $dT_\mathrm{min}=0.01$ Gyr, $\alpha=-2.0$ &  0.005 & 0.406 &  ... \\
 $\mathcal{R}_\mathrm{12}$ &  GRB: \cite{Salvaterra2012} &  PL, $dT_\mathrm{min}=0.01$ Gyr, $\alpha=-2.0$ &  0.000 & 0.357 &  ... \\
 $\mathcal{R}_\mathrm{13}$ &  GRB: \cite{Palmerio2021} &  PL, $dT_\mathrm{min}=0.01$ Gyr, $\alpha=-2.0$ &  \it{0.014} & 0.417 &  ... \\
 $\mathcal{R}_\mathrm{14}$ &  SFH: \cite{Li2008} &  PL, $dT_\mathrm{min}=5.0$ Gyr, $\alpha=-1.0$ &  \bf{0.799} & 0.517 &  ... \\
 $\mathcal{R}_\mathrm{15}$ &  GRB: \cite{Lien2014} &  PL, $dT_\mathrm{min}=5.0$ Gyr, $\alpha=-1.0$ &  \bf{0.144} & 0.463 &  \ref{fig:Rgrb_best} \\
 $\mathcal{R}_\mathrm{16}$ &  GRB: \cite{Salvaterra2012} &  PL, $dT_\mathrm{min}=5.0$ Gyr, $\alpha=-1.0$ &  \bf{0.325} & 0.476 &  ... \\
 $\mathcal{R}_\mathrm{17}$ &  GRB: \cite{Palmerio2021} &  PL, $dT_\mathrm{min}=5.0$ Gyr, $\alpha=-1.0$ &  \bf{0.676} & 0.524 &  ... \\
 $\mathcal{R}_\mathrm{18}$ &  SFH: \cite{Li2008} &  Cst Delay, $\tau=6.5$ Gyr &  \bf{0.533} & 0.494 &  \ref{fig:models_comparison} \\
 $\mathcal{R}_\mathrm{19}$ &  GRB: \cite{Lien2014} &  Cst Delay, $\tau=7.0$ Gyr &  \it{0.014} & 0.422 &  ... \\
 $\mathcal{R}_\mathrm{20}$ &  GRB: \cite{Salvaterra2012} &  Cst Delay, $\tau=7.0$ Gyr &  \it{0.035} & 0.458 &  \ref{fig:Rgrb_best} \\
 $\mathcal{R}_\mathrm{21}$ &  GRB: \cite{Palmerio2021} &  Cst Delay, $\tau=6.5$ Gyr &  \bf{0.255} & 0.502 &  \ref{fig:Rgrb_d}, \ref{fig:Rgrb_best} \\
 $\mathcal{R}_\mathrm{22}$ &  SFH: \cite{Li2008} &  logNorm, $\mathrm{t_d}=7.3$ Gyr, $\sigma_t = 0.3$ &  \bf{0.966} & 0.499 &  ... \\
 $\mathcal{R}_\mathrm{23}$ &  GRB: \cite{Lien2014} &  logNorm, $\mathrm{t_d}=10.0$ Gyr, $\sigma_t = 0.3$ &  \bf{0.892} & 0.519 &  \ref{fig:Rgrb_best} \\
 $\mathcal{R}_\mathrm{24}$ &  GRB: \cite{Salvaterra2012} &  logNorm, $\mathrm{t_d}=8.6$ Gyr, $\sigma_t = 0.3$ &  \bf{0.990} & 0.498 &  ... \\
 $\mathcal{R}_\mathrm{25}$ &  GRB: \cite{Palmerio2021} &  logNorm, $\mathrm{t_d}=7.3$ Gyr, $\sigma_t = 0.3$ &  \bf{0.986} & 0.504 &  \ref{fig:Rgrb_d}, \ref{fig:Rgrb_best}, \ref{fig:models_comparison} \\
 $\mathcal{R}_\mathrm{26}$ &  SFH: \cite{Li2008} &  logNorm, $\mathrm{t_d}=5.0$ Gyr, $\sigma_t = 1.0$ &  \it{0.023} & 0.437 &  ... \\
 $\mathcal{R}_\mathrm{27}$ &  GRB: \cite{Lien2014} &  logNorm, $\mathrm{t_d}=5.0$ Gyr, $\sigma_t = 1.0$ &  \it{0.012} & 0.424 &  ... \\
 $\mathcal{R}_\mathrm{28}$ &  GRB: \cite{Salvaterra2012} &  logNorm, $\mathrm{t_d}=5.0$ Gyr, $\sigma_t = 1.0$ &  0.007 & 0.416 &  ... \\
 $\mathcal{R}_\mathrm{29}$ &  GRB: \cite{Palmerio2021} &  logNorm, $\mathrm{t_d}=5.0$ Gyr, $\sigma_t = 1.0$ &  \it{0.017} & 0.436 &  ... \\
\bottomrule
\end{tabular}
}
\end{table*}

The values of $\langle \mathrm{N/N_{max}} \rangle$, the mean of the \nnmax\ distribution, and the p-values of the KS test are given in Table \ref{tab:ks_results} for selected models. The first 5 models are a constant density rate evolution in addition to the SFH and the three GRB models presented in the methodology section, without any delay. They are followed by 12 models with a power-law delay function, with two values of the power-law index, $\alpha = -2$, and $\alpha = -1$, and two values of $dT_\mathrm{min}$, $0.01$ and $5$ Gyr. The first value of $dT_\mathrm{min}$ is standard for such models, while larger values of this parameter have been used to explore its impact on the rate density distributions. Finally, we also consider 12 models with a lognormal delay function, with three values of the dispersion: $\sigma_\mathrm{t} = 0.01$, $0.3$ and $1.0$, and for each of them we indicate the delay $t_\mathrm{d}$ that gives the largest p-value. As explained before, the model with $\sigma_\mathrm{t} = 0.01$ is equivalent to a constant delay model.

According to Table \ref{tab:ks_results}, ten models have p-values larger than 10\% (favored models), while 11 models have a p-value between 1 and 10\%. All favored models are delayed models, with a significant delay between the GRB and the merger event:
\begin{itemize}

\item The SFH model and the GRB models of \citeauthor{Salvaterra2012}, \citeauthor{Lien2014} and \citeauthor{Palmerio2021} convolved with a power-law delay function with $\alpha=-1$ and $dT_\mathrm{min}=5$ Gyr ($\mathcal{R}_{14}$, $\mathcal{R}_{15}$, $\mathcal{R}_{16}$, $\mathcal{R}_{17}$).

\item The SFH model and the GRB model of \citeauthor{Palmerio2021} convolved with a constant delay function of \mbox{7 Gyr} ($\mathcal{R}_{18}$, $\mathcal{R}_{21}$).

\item The SFH model and the GRB models of \citeauthor{Lien2014}, \citeauthor{Salvaterra2012} and \citeauthor{Palmerio2021} convolved with a lognormal delay function with $\sigma_\mathrm{t}=0.1$ or $0.3$ and optimal $t_\mathrm{d}$ values ($\mathcal{R}_{22}$, $\mathcal{R}_{23}$, $\mathcal{R}_{24}$, $\mathcal{R}_{25}$).

\end{itemize}

Some favored models are shown in Fig. \ref{fig:Rgrb_z}c, for instance the model of \citeauthor{Lien2014} with a delay $t_\mathrm{d} = 10$ Gyr and a dispersion $\sigma_\mathrm{t}=0.3$ ($\mathcal{R}_{23}$). A common feature of these models is the decreasing density of mergers beyond z $\approx 0.8$. This is due to the fact that the temporal delay between the GRB and the merger transforms the peak of GRB production (at \mbox{z $\approx 2-3$}) into a peak of merger production at \mbox{z $\approx 0.5$}.

Regarding models which are not-favored (rejected or simply acceptable), they all have values of \mbox{$<\mathrm{N/N_{max}}>$} below $0.5$, indicating that they predict too many distant BBH mergers, compared with GW observations. This is well illustrated by the cumulative distribution function of \nnmax\ for models $\mathcal{R}_4$ or $\mathcal{R}_{11}$ in Fig. \ref{fig:models_comparison}. 

\begin{figure}
\gridline{\fig{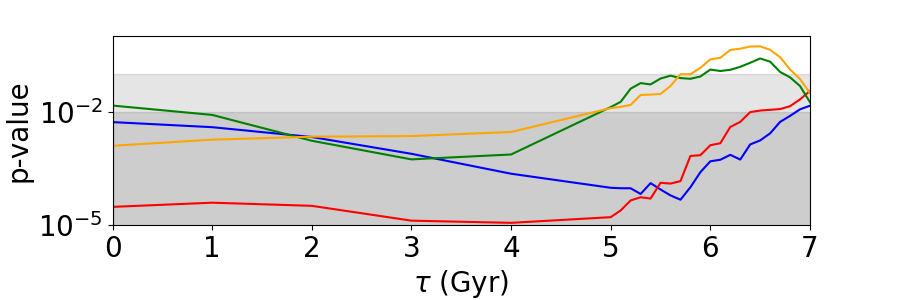}{0.45\textwidth}{\label{fig:cst_pvalue}(a) Constant delay function}}
\gridline{\fig{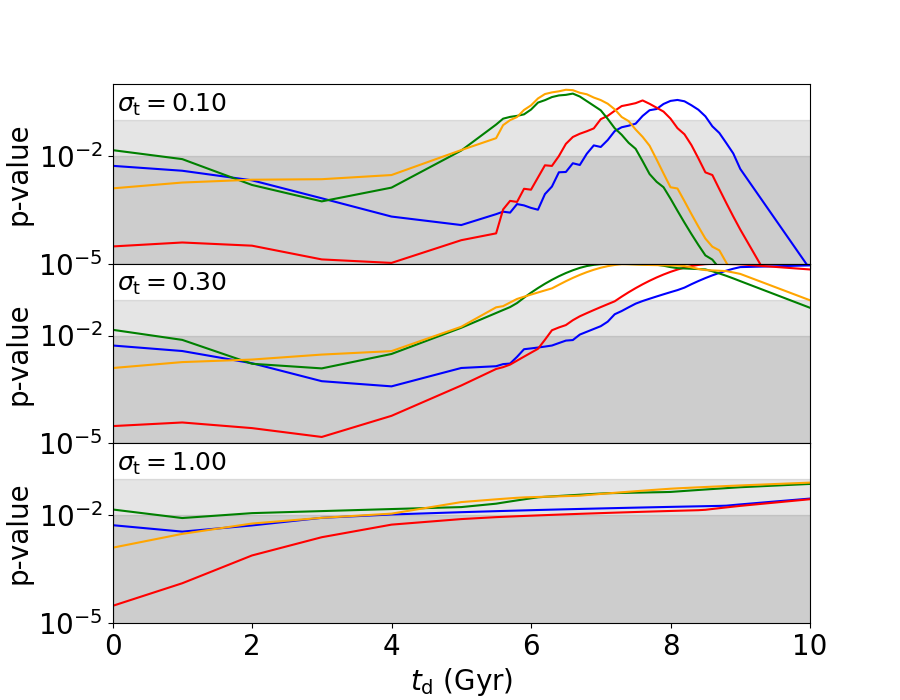}{0.45\textwidth}{\label{fig:lognormal_pvalue}(b) Lognormal delay function}}
\gridline{\fig{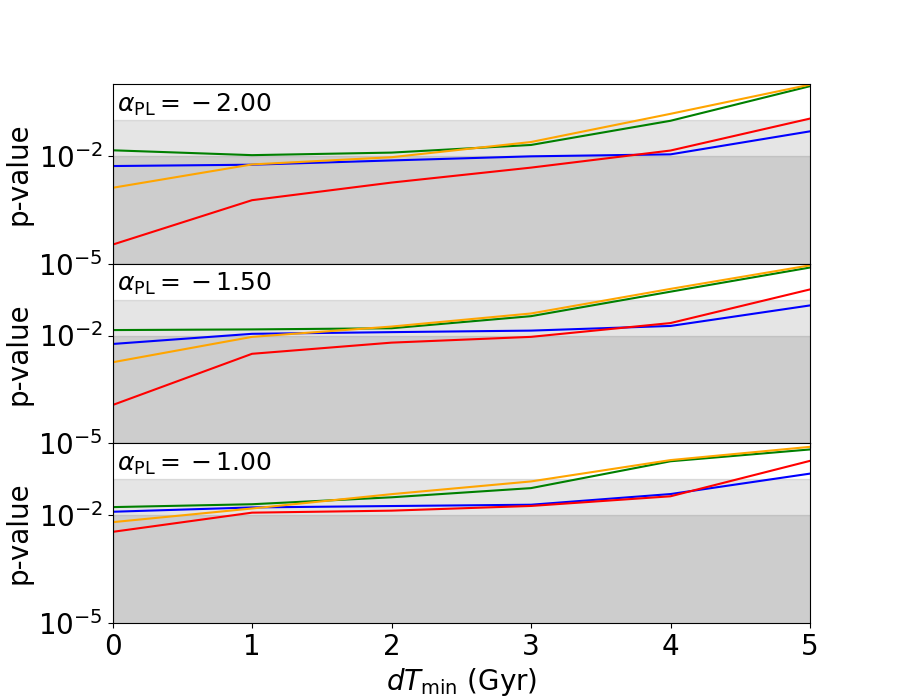}{0.45\textwidth}{\label{fig:dTmin_pvalue}(c) Power-Law delay function}}
\caption{Evolution of the p-value with the delay, for various delay functions applied to the three GRB world models considered in the paper (blue, red and green for respectively \citealt{Lien2014}, \citealt{Salvaterra2012} and \citealt{Palmerio2021}) and the SFH from \cite{Li2008} in orange. The shaded light-grey and dark-grey areas represent the thresholds p-values for the accepted and favored models, respectively set to 1\% and 10\%.}
\label{fig:delay_pvalue}
\end{figure}

Figure \ref{fig:delay_pvalue} provides an insight into the impact of the delay for the three delay functions studied here, it emphasizes some trends: (\textit{i}) SFH or GRB populations with no or short delay are not favored; (\textit{ii}) When a delay is included, the three delay functions can reproduce the observed distribution if the delay is sufficiently large: larger than \mbox{$\sim 6$ Gyr} for the constant and lognormal delay functions, and $dT_\mathrm{min}$ larger than \mbox{$\sim 4$ Gyr} for the power-law delay function; (\textit{iii}) Delays with a broad distribution (e.g. a lognormal distribution with $\sigma_\mathrm{t} = 1$ or a power-law distribution with $dT_\mathrm{min}$ smaller than \mbox{4 Gyr}) are not favored, and finally, (\textit{iv}) the crucial parameter for the power-law distribution is $dT_\mathrm{min}$, the slope playing a marginal role. 

\subsection{Impact of systematics}
\label{sub:systematics}

These conclusions rely on the accuracy of the redshift measurement by LVC and on the calculation of the merger horizon. 
\begin{figure}
\resizebox{\hsize}{!}{\includegraphics{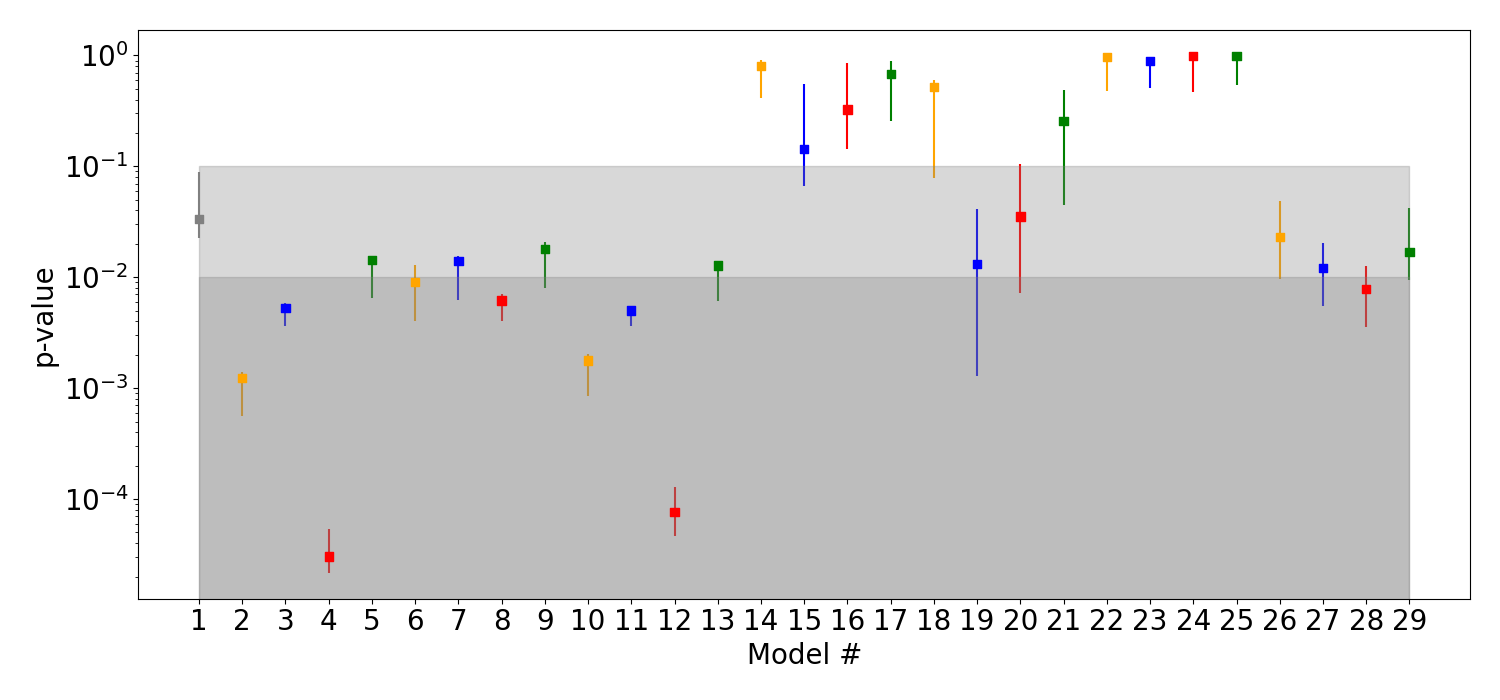}}
\caption{Effect of the redshift uncertainties on the p-values obtained for the selected models. The range displayed is the 90\% confidence interval on the p-value. The shaded grey area represents the threshold below which models are rejected, set to \alphap}
\label{fig:error_z}
\end{figure}

The effect of redshift uncertainties on the p-values has been investigated using the redshift limits given in GWTC-1, GWTC-2 and GWTC-3. The redshifts of BBH mergers are poorly constrained in these catalogs, with a relative median redshift uncertainty of 39\%. We have randomized the redshift values given in the catalog (assuming that the incertitude on the redshift could be modeled by two half-Gaussian distributions whose standard deviations are equal to the 90\% confidence interval divided by 1.65) and tested $10,000$ times the p-value obtained for each model. Figure \ref{fig:error_z} shows the 90\% confidence interval obtained for the p-values of a given model. 
When the redshift uncertainty is taken into account, the accepted models $\mathcal{R}_{5}$ $\mathcal{R}_{7}$, $\mathcal{R}_{9}$, $\mathcal{R}_{13}$, $\mathcal{R}_{19}$, $\mathcal{R}_{20}$, $\mathcal{R}_{27}$ and $\mathcal{R}_{29}$ can be rejected for some cases of redshift, while models $\mathcal{R}_{6}$ and $\mathcal{R}_{28}$ become acceptable. These models are referred as marginally acceptable in the remaining of the paper.
The list of favored models does not change, except for model $\mathcal{R}_{15}$, $\mathcal{R}_{18}$ and $\mathcal{R}_{21}$ that becomes only acceptable for some cases. These models are referred as marginally favored in the remaining of the paper.

We conclude that our study is marginally sensitive to statistical uncertainties on the measured merger redshifts. The findings of this article, discussed in the next section, are mainly limited by the number of BBH mergers detected until now.

\section{The rate of BBHs} 
\label{sec:rate}

In this section, we evaluate the rate density of BBH merger progenitors necessary to produce the BBH mergers density rate observed by LVC, considering favored and marginally favored models identified in Sec. \ref{sec:distriz} (with a p-value larger than 10\%). This density rate will be expressed in terms of the LGRBs density rate, the models using the SFH density rate evolution are therefore not considered here. Using the LGRB density rate as a reference does not imply a connection between LGRBs and BBH mergers, this is just a convenient way of specifying the density rate. We defer to section \ref{sec:discussion} the discussion of the possible connections between the two populations.

\subsection{Methodology}

Following the approach of Equ. \ref{equ:Rbbhs_z}, we write the expected rate of BBHs mergers $\rho(z_0)$ as a function of their birth rate  $\mathcal{R}_\mathrm{BBHs}(z)$ and the normalized delay distribution $f(z,z_0)$: 
\begin{equation}
    \rho(z_0) = \int_{\mathrm{z_0}}^{\infty} \mathcal{R}_\mathrm{BBHs}(z)\ f(z, z_0)\frac{\mathrm{d}T_c}{\mathrm{d}z}\mathrm{d} z
\end{equation}
The best measurement of the BBH merger rate density in GWTC-3 \citep{GWTC3_population} occurs at $z=0.2$. For this reason, the equation above will be computed at $z_0=0.2$ to get an estimation of the fraction as precise as possible. Using the LGRB rate as a reference, the BBH birth rate $\mathcal{R}_\mathrm{BBHs}(z)$ can be linked to the observed LGRBs rate using two factors, $f_\mathrm{b}$ and $\eta_0$:

\begin{equation}
    \mathcal{R}_\mathrm{BBHs}(z) = \eta_0 \times f_\mathrm{b} \times \mathcal{R}_\mathrm{GRB}(z)
\end{equation}

The beaming factor $f_\mathrm{b}$ represents the ratio of total number of LGRBs to the number of \textit{detected} LGRBs pointing towards us. It is sometimes expressed relatively to the jet opening angle $\theta$, as $f_\mathrm{b} = (1-\cos \theta )^{-1}$. It is assumed to be independent of redshift.\\

The parameter $\eta_0$ is a normalization factor to be applied to LGRBs density rate to reproduce the observed rate of BBH mergers. 
$\eta_0$ is the number that we aim to constrain in this section, as it permits to compare the relative rates of BBH mergers and LGRBs.
$\eta_0$ has been assumed constant and independent of redshift in this analysis. However, it is also possible to assume a difference in evolution between massive BBHs and GRBs, favoring or disfavoring the production of BBH systems relatively to the LGRBs. This evolution can follow a simple parametrization of the density population (e.g. $\eta(z) = \eta_0 (1+z)^{0.6}$, similarly as in \citealt{Salvaterra2012}). It can also be assumed that the BBHs production is favored at low metallicity, meaning that the evolution of $\eta(z)$ is inversely proportional to the metallicity evolution ($\eta(z)= \eta_0 10^{0.15z}$, using the metallicity evolution from \citealt{Li2008}). In both models, this factor boosts the BBHs density evolution by a factor $\times2$ at $z=2$ compared to LGRBs. 
\\ 

Given the two equations above, it is possible to compute the fraction $\eta_0$ as:

\begin{equation}
    \eta_0 = \frac{\rho(z_0)}{f_\mathrm{b}} \times \left(  \int_{z_0}^{\infty} \mathcal{R}_\mathrm{GRB}(z)\ f(z, z_0)\frac{\mathrm{d}T_c}{\mathrm{d}z}\mathrm{d} z \right)^{-1}
\label{equ:ratio}
\end{equation}

\subsection{Results}

\begin{figure}
\centering
\resizebox{\hsize}{!}{\includegraphics{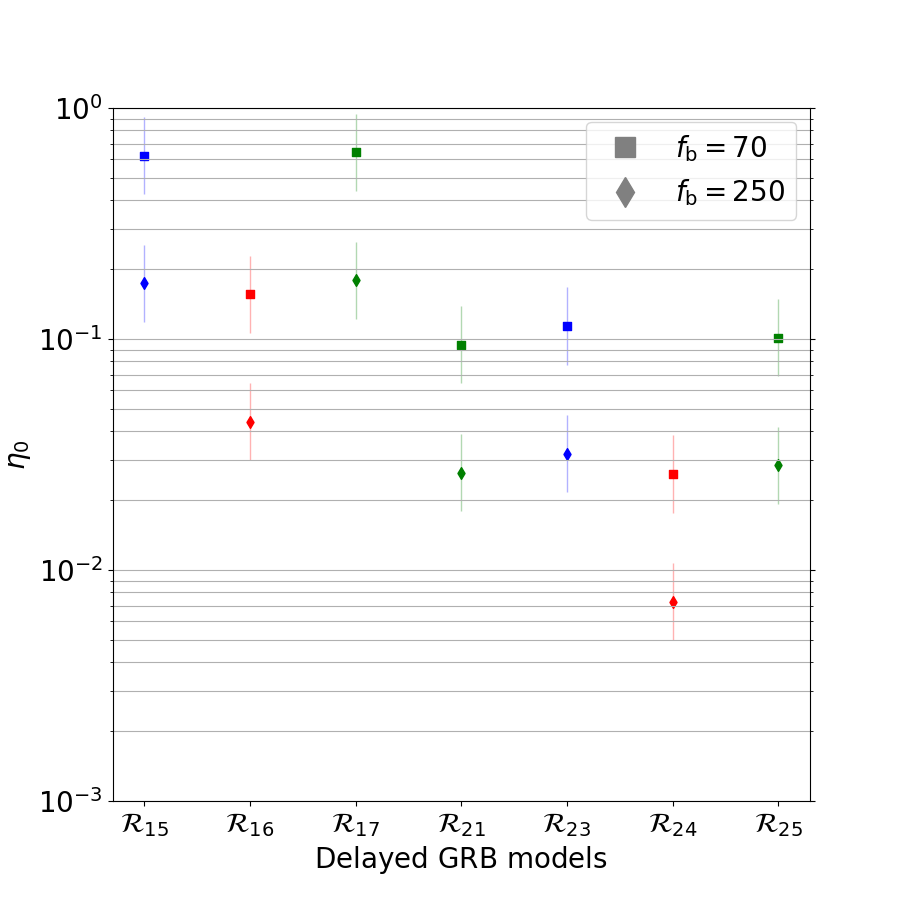}}
\caption{The rate of BBH mergers in units of the LGRB rate, for the 7 favored GRB models. The error reflects the incertitude on the value of $\rho(z_0=0.2)$. The marker shapes represent combinations of beaming factors ($f_\mathrm{b}=70, 250$).}
\label{fig:eta0}
\end{figure}

Figure \ref{fig:eta0} shows the values of $\eta_0$ calculated for the 7 favored models in the previous section, for two values of the GRB beaming factor, $f_\mathrm{b} = 250$ and $f_\mathrm{b} = 70$, respectively corresponding to jet opening angles of about $5$ and $9$ degrees.
In the rest of the discussion, we consider $f_\mathrm{b} = 250$ as a typical value \citep{Goldstein2016, Tsvetkova2017, Lamb2021}, while $f_\mathrm{b} = 70$ is to be considered as a lower bound to the beaming factor.
For the BBH mergers rate $\rho(z_0)$, we adopt the value given in the GWTC-3 catalog \citep{GWTC3_population},  $\rho(z_0=0.2) = 19-41~\mathrm{Gpc}^{-3}~\mathrm{yr}^{-1}$.
Taking into account the uncertainties on LVC BBH mergers local density rate, equation \ref{equ:ratio} gives a large range of predictions for $\eta_0$, ranging from $\sim 0.01$ to $\sim 1$ (see Fig. \ref{fig:eta0}).
Considering the beaming factor $f_\mathrm{b}=250$, the value of $\eta_0$ is typically of few percents for models $\mathcal{R}_{16}$, $\mathcal{R}_{21}$, $\mathcal{R}_{23}$ and $\mathcal{R}_{25}$, going up to ten percents for models $\mathcal{R}_{15}$ and $\mathcal{R}_{17}$ and as low as few tenths of a percent for model $\mathcal{R}_{24}$. The values for a beaming factor $f_\mathrm{b}=70$ are about three times larger.
We note that for models $\mathcal{R}_{15}$, $\mathcal{R}_{16}$ and $\mathcal{R}_{17}$, the delay between the GRB and the merger follows a power-law with DTmin = 5 Gyr, meaning that there is no merger during 5~Gyr after a LGRB. These models, while acceptable from a statistical point of view, seem little realistic from an astrophysical point of view.\\
Figure \ref{fig:eta0} also shows that this ratio is below $1$ for all acceptable models and $f_\mathrm{b}$ larger than 70. If BBH mergers and LGRBs come from the same BH population, BBH mergers must be rarer than LGRBs.
Considering $f_\mathrm{b} = 250$ as previously justified, BHs in BBH mergers may represent at most a few percent of those produced in LGRBs. The vast majority of BHs produced in LGRBs do not end their lives in BBH mergers.

\section{Discussion}
\label{sec:discussion}
\subsection{Constraints on the BBH merger population}
\label{sub:constraints}

Section \ref{sec:distriz} shows that BBH mergers and LGRBs cannot have the same formation history, unless BBH mergers have a long coalescence time of several Gyr. We briefly discuss this assumption here.
Since the LGRB formation rate peaks about 2-3 Gyr after the big bang, models that consider a delay of $t_\mathrm{d}$ Gyr between the GRB and the merger predict a peak of mergers at 3+$t_\mathrm{d}$ Gyr after the big bang. For values of $t_\mathrm{d}$ smaller than 5~Gyr, this peak occurs beyond the horizon of GW interferometers, leading to a rate of BBHs that steadily increases with the redshift in the range of current GW interferometers ; such models are not compatible with the data. On the other hand, models with a longer delay ($t_\mathrm{d} \approx 5-6$~Gyr), which lead to a peak of BBH mergers formation at redshift $z \approx 0.5 - 0.6$, appear favored by the data. These models imply a decrease of the BBH mergers beyond the peak (Fig. \ref{fig:Rgrb_z}c), providing a direct mean to test the assumption that BBH mergers and LGRBs have progenitors with the same formation history. \\
Most evolutionary models of BBH mergers, however, predict a delay function following a power-law with steep index $\alpha$ and low $dT_\mathrm{min}$, with a mean delay time close to $\sim 1\ \mathrm{Gyr}$ \citep[e.g.][]{Dominik2012}, even if other studies suggest that a delay time of several Gyr might also be possible for BH-BH systems \citep{Peters1964, Abbott2016, Broekgaarden2021}.
Since coalescence times of several Gyr are hardly credible, we conclude that BBH mergers and LGRBs do not have the same formation history.
This conclusion needs, of course, to be confirmed with larger BBH merger samples, which will become available in a near future, with runs O4 and O5.

In Sec. \ref{sec:rate}, we compare the rate of BBH mergers measured by the LVC collaboration with the rate of LGRBs. Assuming again that BBH mergers and LGRBs have the same formation history, we show that BBH mergers are rarer than LGRBs: taking for example the model $\mathcal{R}_{25}$ and a standard value for the beaming factor $f_\mathrm{b}=250$, the normalization factor $\eta_0$ is between $2$ and $4$\%. If the two populations share the same formation history, several percent of the LGRBs suffice to explain the measured rate of BBH mergers, and if not, this is an indication of a lack of connection between LGRBs and BBH mergers. In both cases, the large majority of the BHs born in LGRBs do not end their lives in BBH mergers. \\
Taken together these two studies suggest that BBH mergers and LGRBs have distinct progenitors, and that LGRBs are more frequent than BBH mergers.

\subsection{The nature of LGRB progenitors}
\label{sub:origin_LGRBs}

 One of the surprises of LVC detections was the dominance of BHs with a low spin. This tells us that the cores of massive stars have lost most of their angular momentum when they collapse into a BH. The core angular momentum has been transferred to the stellar envelope which has been ejected. This is the indication of an efficient transport of angular momentum in the progenitors of BBH mergers and possibly in most massive stars \citep{Olejak2021, Bavera2021, Fuller2022}. If this scenario is correct, LGRBs, which require stellar cores with a large angular momentum at BH creation, cannot be emitted by single massive stars, nor by massive stars in wide binary systems. According to \citet{Kushnir2017, Piran2020, Bavera2020, Belczynski2020, Marchant2021} the best way for a massive stellar core to keep a large angular momentum until it becomes a BH is through tidal spin-up in a compact system, which is bound to merge after some time. In this model, the population of LGRBs and BBHs  (or at least a sub-sample of the latter) would therefore share common progenitors.

 \cite{Bavera2021} have recently published a study using binary stellar evolution and population synthesis calculations to explain the sub-population of spinning merging BBHs and the emission of LGRBs at the BH formation. They reach the conclusion that the progenitors of fast-spinning BBH mergers, formed via isolated binary evolution, are likely a major contribution to the observed rate of luminous LGRBs. 
 
 This conclusion differs from the results obtained in this paper, as we propose different origins for BBH mergers and LGRBs. This tension can be explained as follow:
\begin{itemize}
    \item The population of BH progenitors generated by \cite{Bavera2021} for the production of LGRBs is based on a value close to the upper limit of the local rate measured by LVC, $\rho(z_0=0) = 38.3~\mathrm{Gpc}^{-3}~\mathrm{yr}^{-1}$. We have taken the logarithmic mean of the upper- and lower- limits ($\rho(z_0=0.2) = 28~\mathrm{Gpc}^{-3}~\mathrm{yr}^{-1}$), which is smaller.
    \item The SHOALS survey used as a reference for the LGRBs distribution in \cite{Bavera2021} gives a local rate of observed GRBs $\sim0.2~\mathrm{Gpc^{-3}~yr^{-1}}$, while the ones used in this papers are larger, and closer to $1~\mathrm{Gpc^{-3}~yr^{-1}}$.
    \item \cite{Bavera2021} use a beaming fraction of $0.05$ (corresponding to a beaming factor of $f_\mathrm{b} = 20$ or a jet opening angle of $\theta_\mathrm{j}\sim 18$°), while we consider a larger beaming factor, by a factor $\sim 3-12$.  \cite{Goldstein2016} show that the distribution of jet opening angles derived from the prompt emission of 638 GRBs observed by the Fermi Gamma-ray Burst Monitor (GBM) peaks at $\sim 6$° or $f_\mathrm{b}=180$, and \cite{Tsvetkova2017} find a median opening angle of $\theta_\mathrm{j}\sim 4$° ($f_\mathrm{b}=400$), based on breaks in the afterglows of 43 bright GRBs detected by KONUS. Independent measurements obtained by \cite{Lamb2021} on long GRB afterglows also seem to point towards a jet opening angle around $\sim 5$° ($f_\mathrm{b}=260$). As the beaming factor is directly connected to the true GRB rate, the different values taken by \cite{Bavera2021} and us explain our different results.
\end{itemize}
These three factors, which all go in the same direction, lead to a ratio of $\sim 100$ between the long GRB rate used by \cite{Bavera2021} and the one we use. 
This explains why we find that at most a small fraction of LGRBs could be associated to BBH mergers, while \cite{Bavera2021} propose to associate all long GRBs with the sub-population of BBH mergers with non-zero spin.
In the rest of this discussion we consider that our LGRB models (including the beaming factor) are realistic, and we discuss our findings in this context.

 If the majority of LGRBs do not end their lives in BBH mergers, they cannot be produced in compact binary systems and must be emitted by single stars \citep[e.g.][]{Aguilera2020} or in binary systems that do not merge.
 It may be natural to associate them with systems like the X-ray Binaries (LMXBs, HMXBs) found in the galaxy, such as Cygnus X-1 \citep{Krawczynski2022, Zhao2021}, GRS 1915+105 \citep{Shreeram2020} or EXO 1846-031 \citep{Draghis2020} which are rapidly spinning and could be the descendants of LGRBs (although \cite{Belczynski2021_all_apples} show that BH spins determination can be subject to systematic effects). The detection of such systems, especially HMXBs for which accretion spin-up is limited by the lifetime of the secondary star \citep[e.g.][]{Wong2012}, requires the birth of highly spinning BHs \citep{Qin2019, MillerJones2021}, an expected characteristics of LGRB descendants.
Nonetheless, it is also possible that LGRB descendants are partitioned in several populations, X-ray binaries being only a subset of them.

\subsection{Short-lived LGRB descendants in the BBH merger population ?}
\label{sub:LGRBs_bbhs}
\begin{figure}
\resizebox{\hsize}{!}{\includegraphics{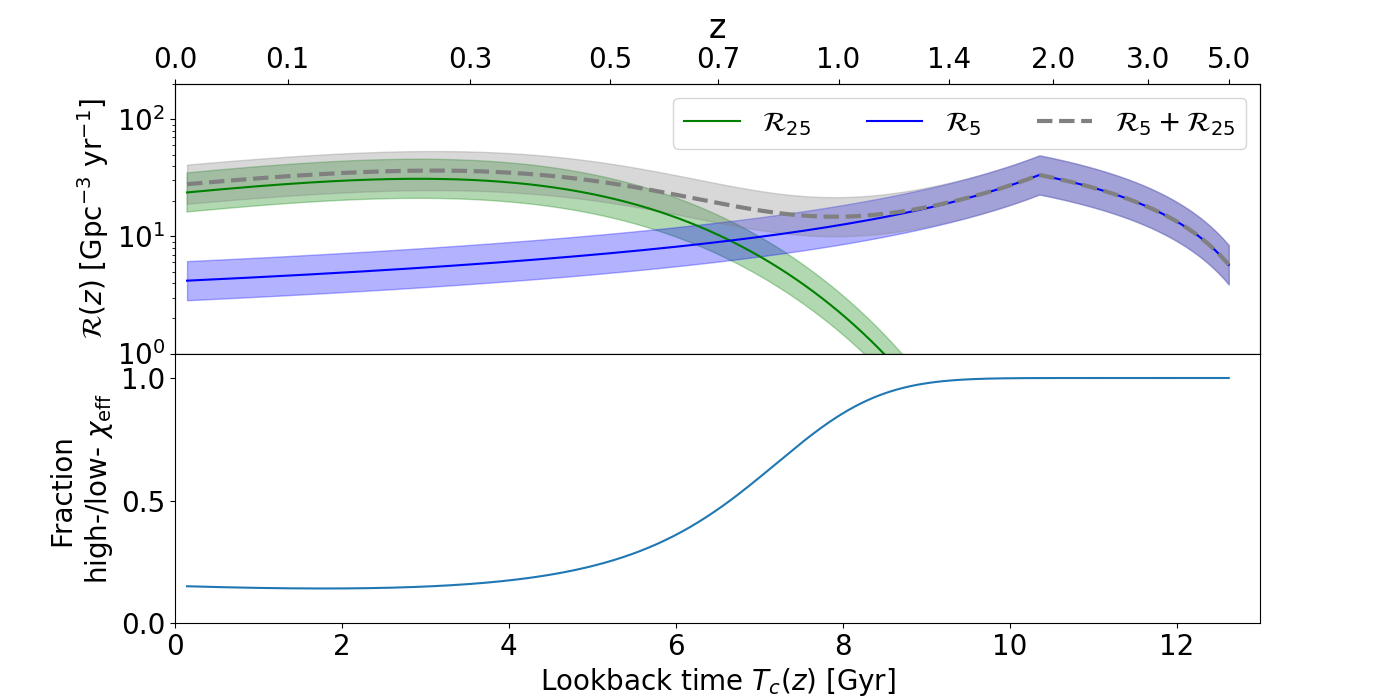}}
\caption{Illustration of the potential two-population: the rate density presented on top is composed at $85$\% by a high-delay (low-spin) population and $15$\% by a low-delay (high-spin) population, the sum of the two illustrated by the dashed grey line. The bottom panel illustrates the evolution of the high-spin fraction of BBHs mergers relatively to redshift.}
\label{fig:two_contributions}
\end{figure}

While our analysis suggests that the majority of BBH mergers have no direct connection with LGRBs, it does not preclude the existence of a small fraction of GRB descendants, with non-zero spin and short merging times, among BBH mergers. Such systems could be produced during a phase of tidal spin-up in a compact system of two massive stars (or a BH and a massive star) as proposed by \citet{Bavera2021}. We show here that the addition of a minority fraction of short-lived GRB descendants to a favored model does not change its acceptability.\\
For illustration we have constructed a mixed model, made of 85\% of sources following model $\mathcal{R}_{25}$ and 15\% of sources following the GRB distribution of \citet{Palmerio2021} (model $\mathcal{R}_{5}$). The fraction used roughly corresponds to the fraction of BBH mergers in our sample with a $\chi_\mathrm{eff}$ consistent with being positive.
This mixed model, shown in Figure \ref{fig:two_contributions}, is acceptable for the \nnmax\ test (p = 0.90). It requires the following fractions of the GRB population to explain the two components of the model: 3\% for sources following the $\mathcal{R}_{25}$ distribution and 2\% for sources following the \citet{Palmerio2021} distribution ($\mathcal{R}_{5}$). 

The addition of a sub-population of GRB descendants with short merging times has an interesting consequence: this population becomes dominant beyond a certain redshift ($z \approx 0.7$ in Fig. \ref{fig:two_contributions}). This is the consequence of the weak evolution of BBH mergers (cf. Sec. \ref{sec:distriz}) compared with the strong evolution of GRBs. This provides a test of the existence of a fraction of direct GRB descendants with short merging times among BBH mergers. If this population exists, it may become dominant at intermediate redshifts and start to be detectable in runs O4 and O5.

Another prediction of having GRBs exploding in compact binary systems is the possibility of precessing black holes and precessing GRB jets, that may leave an imprint on the prompt emission or the afterglow of some LGRBs. This possibility has been discussed by various authors, in different contexts \citep{Blackman1996, Fargion2006, Huang2021}.

\section{Conclusion and perspectives}
\label{sec:conclusion}

With the increasing number of stellar mass BH detections, population studies become effective for the comparison of their various sub-classes. Section \ref{sec:distriz} shows the comparison between the BBH merger population and a BH population traced by LGRBs, showing that their redshift distributions are compatible only if we consider a delay of several Gyr between the occurrence of the LGRB and the BBH merger. Section \ref{sec:rate} shows that for most of the compatible models, the BBH merger progenitors rate density represents less than $10$\% of the LGRBs rate density. These two results are interpreted in Sec. \ref{sub:constraints} as an indication that BBH mergers and LGRBs have distinct progenitors, with the proposal that the descendants of LGRBs could resemble more BHs found in galactic X-ray binaries than those of BBH mergers (Sec. \ref{sub:origin_LGRBs}).

After the publication of GWTC-2, various authors have underlined the fact that BBH mergers may not be a single population \citep[e.g.][]{Bouffanais2021, Hutsi2021, Zevin2021}, but a mix of BH populations originating from isolated stellar binaries, dynamic mergers, and possibly primordial BHs. This claim, however, has been challenged by other authors \citep[e.g.][]{Belczynski2021_uncertain, Broekgaarden2021}. The same is true for LGRBs, which can have different ``engines'', a black hole or a fast rotating magnetar \citep[e.g.][]{Bernardini2015}, and which encompass various sub-classes like X-ray Flashes, low-luminosity GRBs, choked GRBs, ultra-LGRBs or LGRBs with a plateau in their afterglow.
Future samples of BBH mergers with better statistics will permit to sort out these sub-populations, which are expected to have different mass, spin and redshift distributions. It will then be possible to correlate GRBs with theoretically or empirically defined BBH merger sub-classes (e.g. BBH mergers with non-zero effective spin, within a given mass range), getting more insight into the physical connections between these phenomena.

In this context, larger samples of BBH mergers and GRBs with a redshift originating from various sub-classes will permit population studies of these sub-classes, providing insight into the zoo of astrophysical BHs. Section \ref{sub:LGRBs_bbhs} has shown for example a test for the existence of a sub-population of BBH mergers that would be descendants from LGRBs. Larger samples of BBH mergers are expected with the increasing sensitivity of GW interferometers \citep{prospectsGW_2018}, and larger samples of GRBs with a redshift will be provided by current and future GRB missions doing fast localizations like \textit{Swift} \citep{Gehrels2004, Barthelmy2005}, \textit{Fermi} \citep{Meegan2009}, \textit{INTEGRAL} \citep{Winkler2003, Lebrun2003, Mereghetti2003}, \textit{GECAM} \citep{Lv2018} or the coming \textit{SVOM} \citep{Wei2016, Godet2014, Arcier2020}.

In parallel with these studies, it is essential to develop models of stellar evolution designed to follow the evolution of massive stars up to the production of BHs and their manifestations like GRBs, and BBH mergers \citep[see for instance][]{Bavera2021}. Only with such models guiding us, we will be able to get all the knowledge that can be extracted from the diversity of BH manifestations that are being revealed by present day instruments.

\begin{acknowledgements}
The authors thank the referee for helpful comments on the evolution of binary systems of massive stars that contributed to improve this paper. The authors are grateful to N. Leroy for helpful discussions regarding the operations of the LIGO and Virgo GW detectors. 
This work is supported by CNES and the Paul Sabatier University.
\end{acknowledgements}

\software{NumPy \citep{Numpy}, SciPy \citep{Scipy}, Matplotlib \citep{Matplotlib}, Astropy \citep{Astropy}, LALSuite \citep{lalsuite}}



\appendix
\section{Horizon redshift calculation}
\label{sec:zh_calculation}

Table \ref{tab:z_tab} contains the BBH mergers properties used to obtain the results presented in this paper.

\begin{longtable}{lcccccc}
\caption{\label{tab:z_tab}BBHs mergers used in this paper from \cite{GWTC1, GWTC2, GWTC3}, with the original SNR and the calculated horizon redshift $z_\mathrm{h}$ thanks to Equation \ref{equ:z_h}.}\\
\hline
Name & $m_1 \left(\mathrm{M_\odot}\right)$ & $m_2 \left(\mathrm{M_\odot}\right)$ & SNR & z & $z_\mathrm{h}$ & $V_\mathrm{max}/V$ \\
\hline
\endfirsthead
\caption{continued.}\\
\hline
Name & $m_1 \left(\mathrm{M_\odot}\right)$ & $m_2 \left(\mathrm{M_\odot}\right)$ & SNR & z & $z_\mathrm{h}$ & $V_\mathrm{max}/V$ \\
\hline
\endhead
\hline
GW200316\_215756$^\mathrm{\blacklozenge}$ &  13.1 &  7.8 &  10.3 &  $0.22^{+0.08}_{-0.08}$ &  $0.28^{+0.10}_{-0.10}$ & 2.05 \\
 GW200311\_115853$^\mathrm{\blacklozenge}$ &  34.2 &  27.7 &  17.8 &  $0.23^{+0.05}_{-0.07}$ &  $0.53^{+0.12}_{-0.17}$ & 9.59 \\
 GW200302\_015811$^\mathrm{\blacklozenge}$ &  37.8 &  20.0 &  10.8 &  $0.28^{+0.16}_{-0.12}$ &  $0.38^{+0.22}_{-0.17}$ & 2.33 \\
 GW200225\_060421$^\mathrm{\blacklozenge}$ &  19.3 &  14.0 &  12.5 &  $0.22^{+0.09}_{-0.10}$ &  $0.35^{+0.15}_{-0.16}$ & 3.55 \\
 GW200224\_222234$^\mathrm{\blacklozenge}$ &  40.0 &  32.5 &  20.0 &  $0.32^{+0.08}_{-0.11}$ &  $0.86^{+0.24}_{-0.33}$ & 12.34 \\
 GW200220\_124850$^\mathrm{\blacklozenge}$ &  38.9 &  27.9 &  8.5 &  $0.66^{+0.36}_{-0.31}$ &  $0.71^{+0.39}_{-0.33}$ & 1.18 \\
 GW200219\_094415$^\mathrm{\blacklozenge}$ &  37.5 &  27.9 &  10.7 &  $0.57^{+0.22}_{-0.22}$ &  $0.79^{+0.32}_{-0.32}$ & 2.19 \\
 GW200216\_220804$^\mathrm{\blacklozenge}$ &  51.0 &  30.0 &  8.1 &  $0.63^{+0.37}_{-0.29}$ &  $0.64^{+0.37}_{-0.29}$ & 1.03 \\
 GW200210\_092254$^\mathrm{\blacklozenge}$ &  24.1 &  2.8 &  8.4 &  $0.19^{+0.08}_{-0.06}$ &  $0.20^{+0.08}_{-0.06}$ & 1.14 \\
 GW200209\_085452$^\mathrm{\blacklozenge}$ &  35.6 &  27.1 &  9.6 &  $0.57^{+0.25}_{-0.26}$ &  $0.70^{+0.32}_{-0.33}$ & 1.64 \\
 GW200208\_130117$^\mathrm{\blacklozenge}$ &  37.8 &  27.4 &  10.8 &  $0.40^{+0.15}_{-0.14}$ &  $0.55^{+0.21}_{-0.20}$ & 2.30 \\
 GW200202\_154313$^\mathrm{\blacklozenge}$ &  10.1 &  7.3 &  10.8 &  $0.09^{+0.03}_{-0.03}$ &  $0.12^{+0.04}_{-0.04}$ & 2.41 \\
 GW200129\_065458$^\mathrm{\blacklozenge}$ &  34.5 &  28.9 &  26.8 &  $0.18^{+0.05}_{-0.07}$ &  $0.64^{+0.20}_{-0.28}$ & 30.53 \\
 GW200128\_022011$^\mathrm{\blacklozenge}$ &  42.2 &  32.6 &  10.6 &  $0.56^{+0.28}_{-0.28}$ &  $0.76^{+0.39}_{-0.39}$ & 2.13 \\
 GW200112\_155838$^\mathrm{\blacklozenge}$ &  35.6 &  28.3 &  19.8 &  $0.24^{+0.07}_{-0.08}$ &  $0.62^{+0.20}_{-0.23}$ & 12.76 \\
 GW191230\_180458$^\mathrm{\blacklozenge}$ &  49.4 &  37.0 &  10.4 &  $0.69^{+0.26}_{-0.27}$ &  $0.92^{+0.35}_{-0.36}$ & 1.95 \\
 GW191222\_033537$^\mathrm{\blacklozenge}$ &  45.1 &  34.7 &  12.5 &  $0.51^{+0.23}_{-0.26}$ &  $0.83^{+0.39}_{-0.44}$ & 3.28 \\
 GW191216\_213338$^\mathrm{\blacklozenge}$ &  12.1 &  7.7 &  18.6 &  $0.07^{+0.02}_{-0.03}$ &  $0.16^{+0.05}_{-0.07}$ & 11.78 \\
 GW191215\_223052$^\mathrm{\blacklozenge}$ &  24.9 &  18.1 &  11.2 &  $0.35^{+0.13}_{-0.14}$ &  $0.50^{+0.19}_{-0.21}$ & 2.56 \\
 GW191204\_171526$^\mathrm{\blacklozenge}$ &  11.9 &  8.2 &  17.5 &  $0.13^{+0.04}_{-0.05}$ &  $0.29^{+0.09}_{-0.11}$ & 9.48 \\
 GW191204\_110529$^\mathrm{\blacklozenge}$ &  27.3 &  19.3 &  8.8 &  $0.34^{+0.25}_{-0.18}$ &  $0.38^{+0.28}_{-0.20}$ & 1.31 \\
 GW191129\_134029$^\mathrm{\blacklozenge}$ &  10.7 &  6.7 &  13.1 &  $0.16^{+0.05}_{-0.06}$ &  $0.26^{+0.08}_{-0.10}$ & 4.11 \\
 GW191127\_050227$^\mathrm{\blacklozenge}$ &  53.0 &  24.0 &  9.2 &  $0.57^{+0.40}_{-0.29}$ &  $0.66^{+0.47}_{-0.34}$ & 1.45 \\
 GW191126\_115259$^\mathrm{\blacklozenge}$ &  12.1 &  8.3 &  8.3 &  $0.30^{+0.12}_{-0.13}$ &  $0.31^{+0.13}_{-0.14}$ & 1.11 \\
 GW191109\_010717$^\mathrm{\blacklozenge}$ &  65.0 &  47.0 &  17.3 &  $0.25^{+0.18}_{-0.12}$ &  $0.55^{+0.39}_{-0.26}$ & 8.22 \\
 GW191105\_143521$^\mathrm{\blacklozenge}$ &  10.7 &  7.7 &  9.7 &  $0.23^{+0.07}_{-0.09}$ &  $0.28^{+0.09}_{-0.11}$ & 1.73 \\
 GW191103\_012549$^\mathrm{\blacklozenge}$ &  11.8 &  7.9 &  8.9 &  $0.20^{+0.09}_{-0.09}$ &  $0.22^{+0.10}_{-0.10}$ & 1.36 \\
 GW190930\_133541$^\mathrm{\blacksquare}$ &  12.3 &  7.8 &  9.8 &  $0.15^{+0.06}_{-0.06}$ &  $0.18^{+0.07}_{-0.07}$ & 1.82 \\
 GW190929\_012149$^\mathrm{\blacksquare}$ &  80.8 &  24.1 &  9.9 &  $0.38^{+0.49}_{-0.17}$ &  $0.47^{+0.59}_{-0.21}$ & 1.74 \\
 GW190926\_050336$^\mathrm{\blacksquare}$ &  39.8 &  23.2 &  9.0 &  $0.62^{+0.40}_{-0.29}$ &  $0.71^{+0.47}_{-0.34}$ & 1.37 \\
 GW190925\_232845$^\mathrm{\blacksquare}$ &  21.2 &  15.6 &  9.9 &  $0.19^{+0.07}_{-0.07}$ &  $0.24^{+0.09}_{-0.09}$ & 1.83 \\
 GW190924\_021846$^\mathrm{\blacksquare}$ &  8.9 &  5.0 &  13.2 &  $0.12^{+0.04}_{-0.04}$ &  $0.20^{+0.07}_{-0.07}$ & 4.23 \\
 GW190916\_200658$^\mathrm{\blacksquare}$ &  44.3 &  23.9 &  8.2 &  $0.71^{+0.46}_{-0.36}$ &  $0.73^{+0.47}_{-0.37}$ & 1.07 \\
 GW190915\_235702$^\mathrm{\blacksquare}$ &  35.3 &  24.4 &  13.1 &  $0.30^{+0.11}_{-0.10}$ &  $0.50^{+0.19}_{-0.18}$ & 3.96 \\
 GW190910\_112807$^\mathrm{\blacksquare}$ &  43.9 &  35.6 &  13.4 &  $0.28^{+0.16}_{-0.10}$ &  $0.48^{+0.29}_{-0.18}$ & 4.27 \\
 GW190909\_114149$^\mathrm{\blacksquare}$ &  45.8 &  28.3 &  9.0 &  $0.62^{+0.41}_{-0.33}$ &  $0.71^{+0.47}_{-0.38}$ & 1.39 \\
 GW190828\_065509$^\mathrm{\blacksquare}$ &  24.1 &  10.2 &  11.1 &  $0.30^{+0.10}_{-0.10}$ &  $0.42^{+0.14}_{-0.14}$ & 2.50 \\
 GW190828\_063405$^\mathrm{\blacksquare}$ &  32.1 &  26.2 &  16.0 &  $0.38^{+0.10}_{-0.15}$ &  $0.81^{+0.24}_{-0.36}$ & 6.76 \\
 GW190814$^\mathrm{\blacksquare}$ &  23.2 &  2.6 &  22.2 &  $0.05^{+0.01}_{-0.01}$ &  $0.14^{+0.02}_{-0.03}$ & 19.29 \\
 GW190805\_211137$^\mathrm{\blacksquare}$ &  48.2 &  32.0 &  8.3 &  $0.82^{+0.48}_{-0.40}$ &  $0.85^{+0.50}_{-0.41}$ & 1.08 \\
 GW190803\_022701$^\mathrm{\blacksquare}$ &  37.3 &  27.3 &  8.6 &  $0.55^{+0.26}_{-0.24}$ &  $0.60^{+0.29}_{-0.26}$ & 1.22 \\
 GW190731\_140936$^\mathrm{\blacksquare}$ &  41.5 &  28.8 &  8.5 &  $0.55^{+0.31}_{-0.26}$ &  $0.58^{+0.33}_{-0.28}$ & 1.16 \\
 GW190728\_064510$^\mathrm{\blacksquare}$ &  12.3 &  8.1 &  13.6 &  $0.18^{+0.05}_{-0.07}$ &  $0.31^{+0.09}_{-0.12}$ & 4.58 \\
 GW190727\_060333$^\mathrm{\blacksquare}$ &  38.0 &  29.4 &  12.3 &  $0.55^{+0.21}_{-0.22}$ &  $0.89^{+0.36}_{-0.38}$ & 3.16 \\
 GW190725\_174728$^\mathrm{\blacksquare}$ &  11.5 &  6.4 &  9.1 &  $0.21^{+0.10}_{-0.09}$ &  $0.24^{+0.11}_{-0.10}$ & 1.43 \\
 GW190720\_000836$^\mathrm{\blacksquare}$ &  13.4 &  7.8 &  11.7 &  $0.16^{+0.12}_{-0.06}$ &  $0.23^{+0.18}_{-0.09}$ & 2.97 \\
 GW190708\_232457$^\mathrm{\blacksquare}$ &  17.6 &  13.2 &  13.1 &  $0.18^{+0.06}_{-0.07}$ &  $0.30^{+0.10}_{-0.12}$ & 4.07 \\
 GW190707\_093326$^\mathrm{\blacksquare}$ &  11.6 &  8.4 &  13.0 &  $0.16^{+0.07}_{-0.07}$ &  $0.26^{+0.12}_{-0.12}$ & 4.00 \\
 GW190706\_222641$^\mathrm{\blacksquare}$ &  67.0 &  38.2 &  12.3 &  $0.71^{+0.32}_{-0.27}$ &  $1.10^{+0.45}_{-0.38}$ & 2.70 \\
 GW190701\_203306$^\mathrm{\blacksquare}$ &  53.9 &  40.8 &  11.6 &  $0.37^{+0.11}_{-0.12}$ &  $0.54^{+0.16}_{-0.18}$ & 2.75 \\
 GW190630\_185205$^\mathrm{\blacksquare}$ &  35.1 &  23.7 &  15.6 &  $0.18^{+0.10}_{-0.07}$ &  $0.36^{+0.21}_{-0.15}$ & 6.81 \\
 GW190620\_030421$^\mathrm{\blacksquare}$ &  57.1 &  35.5 &  10.9 &  $0.49^{+0.23}_{-0.20}$ &  $0.68^{+0.32}_{-0.28}$ & 2.28 \\
 GW190602\_175927$^\mathrm{\blacksquare}$ &  69.1 &  47.8 &  12.1 &  $0.47^{+0.25}_{-0.17}$ &  $0.71^{+0.36}_{-0.25}$ & 2.86 \\
 GW190527\_092055$^\mathrm{\blacksquare}$ &  36.5 &  22.6 &  8.9 &  $0.44^{+0.34}_{-0.20}$ &  $0.49^{+0.39}_{-0.23}$ & 1.33 \\
 GW190521\_074359$^\mathrm{\blacksquare}$ &  42.2 &  32.8 &  24.4 &  $0.24^{+0.07}_{-0.10}$ &  $0.78^{+0.25}_{-0.36}$ & 22.02 \\
 GW190521$^\mathrm{\blacksquare}$ &  95.3 &  69.0 &  14.4 &  $0.64^{+0.28}_{-0.28}$ &  $1.08^{+0.40}_{-0.40}$ & 3.36 \\
 GW190519\_153544$^\mathrm{\blacksquare}$ &  66.0 &  40.5 &  12.0 &  $0.44^{+0.25}_{-0.14}$ &  $0.67^{+0.37}_{-0.21}$ & 2.90 \\
 GW190517\_055101$^\mathrm{\blacksquare}$ &  37.4 &  25.3 &  10.2 &  $0.34^{+0.24}_{-0.14}$ &  $0.44^{+0.32}_{-0.19}$ & 2.00 \\
 GW190514\_065416$^\mathrm{\blacksquare}$ &  39.0 &  28.4 &  8.3 &  $0.67^{+0.33}_{-0.31}$ &  $0.70^{+0.35}_{-0.33}$ & 1.10 \\
 GW190513\_205428$^\mathrm{\blacksquare}$ &  35.7 &  18.0 &  12.3 &  $0.37^{+0.13}_{-0.13}$ &  $0.58^{+0.22}_{-0.22}$ & 3.25 \\
 GW190512\_180714$^\mathrm{\blacksquare}$ &  23.3 &  12.6 &  12.3 &  $0.27^{+0.09}_{-0.10}$ &  $0.42^{+0.14}_{-0.16}$ & 3.31 \\
 GW190503\_185404$^\mathrm{\blacksquare}$ &  43.3 &  28.4 &  12.1 &  $0.27^{+0.11}_{-0.11}$ &  $0.41^{+0.17}_{-0.17}$ & 3.21 \\
 GW190426\_190642$^\mathrm{\blacksquare}$ &  106.9 &  76.6 &  9.6 &  $0.70^{+0.41}_{-0.30}$ &  $0.82^{+0.46}_{-0.34}$ & 1.45 \\
 GW190424\_180648$^\mathrm{\blacksquare}$ &  40.5 &  31.8 &  10.1 &  $0.39^{+0.23}_{-0.19}$ &  $0.50^{+0.30}_{-0.25}$ & 1.91 \\
 GW190421\_213856$^\mathrm{\blacksquare}$ &  41.3 &  31.9 &  10.6 &  $0.49^{+0.19}_{-0.21}$ &  $0.66^{+0.27}_{-0.29}$ & 2.12 \\
 GW190413\_134308$^\mathrm{\blacksquare}$ &  47.5 &  31.8 &  9.0 &  $0.71^{+0.31}_{-0.30}$ &  $0.81^{+0.36}_{-0.34}$ & 1.37 \\
 GW190413\_052954$^\mathrm{\blacksquare}$ &  34.7 &  23.7 &  8.6 &  $0.59^{+0.29}_{-0.24}$ &  $0.63^{+0.32}_{-0.26}$ & 1.20 \\
 GW190412$^\mathrm{\blacksquare}$ &  30.1 &  8.3 &  18.9 &  $0.15^{+0.03}_{-0.03}$ &  $0.35^{+0.07}_{-0.07}$ & 11.12 \\
 GW190408\_181802$^\mathrm{\blacksquare}$ &  24.6 &  18.4 &  14.7 &  $0.29^{+0.06}_{-0.10}$ &  $0.55^{+0.12}_{-0.20}$ & 5.48 \\
 GW170823$^\mathrm{\mathbf{O}}$ &  39.5 &  29.0 &  11.5 &  $0.35^{+0.15}_{-0.15}$ &  $0.51^{+0.23}_{-0.23}$ & 2.75 \\
 GW170818$^\mathrm{\mathbf{O}}$ &  35.4 &  26.7 &  11.3 &  $0.21^{+0.07}_{-0.07}$ &  $0.30^{+0.10}_{-0.10}$ & 2.69 \\
 GW170814$^\mathrm{\mathbf{O}}$ &  30.6 &  25.2 &  15.9 &  $0.12^{+0.03}_{-0.04}$ &  $0.24^{+0.06}_{-0.08}$ & 7.36 \\
 GW170809$^\mathrm{\mathbf{O}}$ &  35.0 &  23.8 &  12.4 &  $0.20^{+0.05}_{-0.07}$ &  $0.31^{+0.08}_{-0.11}$ & 3.51 \\
 GW170729$^\mathrm{\mathbf{O}}$ &  50.2 &  34.0 &  10.2 &  $0.49^{+0.19}_{-0.21}$ &  $0.63^{+0.25}_{-0.28}$ & 1.92 \\
 GW170608$^\mathrm{\mathbf{O}}$ &  11.0 &  7.6 &  14.9 &  $0.07^{+0.02}_{-0.02}$ &  $0.13^{+0.04}_{-0.04}$ & 6.20 \\
 GW170104$^\mathrm{\mathbf{O}}$ &  30.8 &  20.0 &  13.0 &  $0.20^{+0.08}_{-0.08}$ &  $0.33^{+0.14}_{-0.14}$ & 4.01 \\
 GW151226$^\mathrm{\mathbf{O}}$ &  13.7 &  7.7 &  13.1 &  $0.09^{+0.04}_{-0.04}$ &  $0.15^{+0.07}_{-0.07}$ & 4.22 \\
 GW151012$^\mathrm{\mathbf{O}}$ &  23.2 &  13.6 &  10.0 &  $0.21^{+0.09}_{-0.09}$ &  $0.26^{+0.11}_{-0.11}$ & 1.89 \\
 GW150914$^\mathrm{\mathbf{O}}$ &  35.6 &  30.6 &  24.4 &  $0.09^{+0.03}_{-0.03}$ &  $0.28^{+0.10}_{-0.10}$ & 25.65 \\

\end{longtable}
\tablecomments{Errors are given at the $90$\% confidence interval. GW events with $^\mathrm{\mathbf{O}}$, $^\mathrm{\blacksquare}$ and $^\mathrm{\blacklozenge}$ are respectively from GWTC-1, GWTC-2 and GWTC-3.}

\bibliography{sample631}{}
\bibliographystyle{aasjournal}



\end{document}